\documentclass[nofootinbib,aps,prc,reprint,superscriptaddress]{revtex4-2}
\usepackage{graphicx}% Include figure files
\usepackage{dcolumn}% Align table columns on decimal point
\usepackage{bm}% bold math
\usepackage{color,soul}% allows highlighting with \hl{}
\usepackage{siunitx} %SI units
\usepackage{xcolor} %additional colours
\usepackage{amsmath} %equation options
\usepackage{amssymb} %Maths symbols
\usepackage{gensymb} %useful for degree o
\usepackage{txfonts}
\usepackage[labelsep=period]{caption}
\usepackage{graphicx} %odd graphics

\usepackage{caption} %adding captions 
\usepackage{subcaption}

\begin{document}

%\preprint{APS/123-QED}

\title{Charge radii, moments and masses of mercury isotopes across the $N=126$ shell closure} 

\author{T.~Day~Goodacre}
\email{tdaygoodacre@triumf.ca}
\affiliation{Department of Physics and Astronomy, School of Natural Science, The University of Manchester, Manchester, M13 9PL, United Kingdom}
\affiliation{CERN, CH-1211 Geneva 23, Switzerland}
\affiliation{TRIUMF, Vancouver V6T 2A3, Canada}
%Collaboration name if desired (requires use of superscriptaddress
%option in \documentclass). \noaffiliation is required (may also be
%used with the \author command).
%\collaboration can be followed by \email, \homepage, \thanks as well.
%\collaboration{}
%\noaffiliation

\author {A.V.~Afanasjev}
\affiliation{Department of Physics and Astronomy, Mississippi State University, MS 39762, USA}

\author {A.E.~Barzakh}
\affiliation{Petersburg Nuclear Physics Institute, NRC Kurchatov Institute, Gatchina 188300, Russia}

\author {L.~Nies}
\affiliation{CERN, CH-1211 Geneva 23, Switzerland}
\affiliation{Universit\"{a}t Greifswald, Institut f\"{u}r Physik, 17487 Greifswald, Germany}

\author {B.A.~Marsh}
\affiliation{CERN, CH-1211 Geneva 23, Switzerland}

\author{S.~Sels}
\affiliation{CERN, CH-1211 Geneva 23, Switzerland}
\affiliation{KU~Leuven, Instituut voor Kern- en Stralingsfysica, B-3001 Leuven, Belgium}

\author {U.C.~Perera}
\affiliation{Department of Physics and Astronomy, Mississippi State University, MS 39762, USA}

\author {P.~ Ring} 
\affiliation{Fakult{\"a}t f{\"u}r Physik, Technische Universit{\"a}t M{\"u}nchen, D-85748 Garching, Germany}

\author{F.~Wienholtz}
\thanks{Present address: Institut f\"{u}r Kernphysik, Technische Universit\"{a}t Darmstadt, 64289 Darmstadt, Germany.}
\affiliation{CERN, CH-1211 Geneva 23, Switzerland}
\affiliation{Universit\"{a}t Greifswald, Institut f\"{u}r Physik, 17487 Greifswald, Germany}

\author {A.N.~Andreyev}
\affiliation{Department of Physics, University of York, York, YO10 5DD, United Kingdom}
\affiliation{Advanced Science Research Center (ASRC), Japan Atomic Energy Agency (JAEA), Tokai-mura, Japan}

\author {P.~Van~Duppen}
\affiliation{KU~Leuven, Instituut voor Kern- en Stralingsfysica, B-3001 Leuven, Belgium}

\author {N.A.~Althubiti}
\affiliation{Department of Physics and Astronomy, School of Natural Science, The University of Manchester, Manchester, M13 9PL, United Kingdom}
\affiliation{Physics Department, Faculty of Science, Jouf University, Aljouf, Saudi Arabia}

\author {B.~Andel}
\affiliation{KU~Leuven, Instituut voor Kern- en Stralingsfysica, B-3001 Leuven, Belgium}
\affiliation{Department of Nuclear Physics and Biophysics, Comenius University in Bratislava, 84248 Bratislava, Slovakia}

\author {D.~Atanasov}
\thanks{Present address: CERN, 1211, Geneva 23, Switzerland.}
\affiliation{Max-Planck-Institut f\"{u}r Kernphysik, Saupfercheckweg 1, 69117 Heidelberg, Germany}

\author {R.S.~Augusto}
\affiliation{TRIUMF, Vancouver V6T 2A3, Canada}

\author {J.~Billowes}
\affiliation{Department of Physics and Astronomy, School of Natural Science, The University of Manchester, Manchester, M13 9PL, United Kingdom}

\author {K.~Blaum}
\affiliation{Max-Planck-Institut f\"{u}r Kernphysik, Saupfercheckweg 1, 69117 Heidelberg, Germany}

\author {T.E.~Cocolios}
\affiliation{Department of Physics and Astronomy, School of Natural Science, The University of Manchester, Manchester, M13 9PL, United Kingdom}
\affiliation{KU~Leuven, Instituut voor Kern- en Stralingsfysica, B-3001 Leuven, Belgium}

\author {J.G.~Cubiss}
\affiliation{Department of Physics, University of York, York, YO10 5DD, United Kingdom}

%\author {A.~de~Roubin}
%\thanks{Present address: Centre d'Etudes Nucl\'{e}aires de Bordeaux-Gradignan, 19 Chemin du Solarium, CS 10120, F-33175 Gradignan, France.}
%\affiliation{Max-Planck-Institut f\"{u}r Kernphysik, Saupfercheckweg 1, 69117 Heidelberg, Germany}

\author {G.J.~Farooq-Smith}
\thanks{Present address: Department of Oncology Physics, Edinburgh Cancer Centre, Western General Hospital, Edinburgh, EH4 2XU, United Kingdom}
\affiliation{Department of Physics and Astronomy, School of Natural Science, The University of Manchester, Manchester, M13 9PL, United Kingdom}
\affiliation{KU~Leuven, Instituut voor Kern- en Stralingsfysica, B-3001 Leuven, Belgium}

\author {D.V.~Fedorov}
\affiliation{Petersburg Nuclear Physics Institute, NRC Kurchatov Institute, Gatchina 188300, Russia}

\author {V.N.~Fedosseev}
\affiliation{CERN, CH-1211 Geneva 23, Switzerland}
%\address[1]{\scriptsize{CERN, CH-1211 Geneva 23, Switzerland}}

%\author {D.A.~Fink}
%\affiliation{Max-Planck-Institut f\"{u}r Kernphysik, Saupfercheckweg 1, 69117 Heidelberg, Germany}
%\affiliation{CERN, CH-1211 Geneva 23, Switzerland}

\author {K.T.~Flanagan}
\affiliation{Department of Physics and Astronomy, School of Natural Science, The University of Manchester, Manchester, M13 9PL, United Kingdom}
\affiliation{The Photon Science Institute, The University of Manchester, Manchester, M13 9PL, United Kingdom}

\author {L.P.~Gaffney}
\thanks{Present address: Department of Physics, University of Liverpool, Liverpool, L69 7ZE, United Kingdom}
\affiliation{KU~Leuven, Instituut voor Kern- en Stralingsfysica, B-3001 Leuven, Belgium}
\affiliation{School of Computing, Engineering, and Physical Sciences, University of the West of Scotland, Paisley PA1 2BE, United Kingdom}

\author {L.~Ghys}
\affiliation{KU~Leuven, Instituut voor Kern- en Stralingsfysica, B-3001 Leuven, Belgium}
\affiliation{Belgian Nuclear Research Center SCK$\bullet$CEN, Boeretang 200, B-2400 Mol, Belgium}

\author {A.~Gottberg}
\affiliation{TRIUMF, Vancouver V6T 2A3, Canada}
\affiliation{University of Victoria, Department of Physics and Astronomy, Victoria, BC V8W 2Y2, Canada}

\author {M.~Huyse}
\affiliation{KU~Leuven, Instituut voor Kern- en Stralingsfysica, B-3001 Leuven, Belgium}

\author {S.~Kreim}
\affiliation{Max-Planck-Institut f\"{u}r Kernphysik, Saupfercheckweg 1, 69117 Heidelberg, Germany}
\affiliation{CERN, CH-1211 Geneva 23, Switzerland}

\author {P.~Kunz}
\affiliation{TRIUMF, Vancouver V6T 2A3, Canada}
\affiliation{Department of Physics, Simon Fraser University, Burnaby, BC, V5A 1S6, Canada}

\author {D.~Lunney}
\thanks{Present address: Universit\'{e} Paris-Saclay, CNRS/IN2P3, IJCLab, 91405 Orsay, France.}
\affiliation{CSNSM-IN2P3, Universit\'{e} de Paris Sud, Orsay, France}

\author {K.M.~Lynch}
\affiliation{Department of Physics and Astronomy, School of Natural Science, The University of Manchester, Manchester, M13 9PL, United Kingdom}
\affiliation{CERN, CH-1211 Geneva 23, Switzerland}

\author {V.~Manea}
\thanks{Present address: Universit\'{e} Paris-Saclay, CNRS/IN2P3, IJCLab, 91405 Orsay, France.}
\affiliation{Max-Planck-Institut f\"{u}r Kernphysik, Saupfercheckweg 1, 69117 Heidelberg, Germany}

\author {Y.~Martinez~Palenzuela}
\affiliation{KU~Leuven, Instituut voor Kern- en Stralingsfysica, B-3001 Leuven, Belgium}
\affiliation{CERN, CH-1211 Geneva 23, Switzerland}

\author {T.M.~Medonca}
\affiliation{CERN, CH-1211 Geneva 23, Switzerland}

\author {P.L.~Molkanov}
\affiliation{Petersburg Nuclear Physics Institute, NRC Kurchatov Institute, Gatchina 188300, Russia}

\author {M.~Mougeot}
\affiliation{CERN, CH-1211 Geneva 23, Switzerland}

\author {J.P.~Ramos}
\affiliation{CERN, CH-1211 Geneva 23, Switzerland}

\author {M.~Rosenbusch}
\thanks{Present address: Wako Nuclear Science Center (WNSC), Institute of Particle and Nuclear Studies (IPNS), High Energy Accelerator Research Organization (KEK), Wako, Saitama 351-0198, Japan}
\affiliation{Universit\"{a}t Greifswald, Institut f\"{u}r Physik, 17487 Greifswald, Germany}

\author {R.E.~Rossel}
\affiliation{CERN, CH-1211 Geneva 23, Switzerland}
\affiliation{Institut f\"{u}r Physik, Johannes Gutenberg-Universit\"{a}t, D-55099 Mainz, Germany}

\author {S.~Rothe}
\affiliation{CERN, CH-1211 Geneva 23, Switzerland}

\author {L.~Schweikhard}
\affiliation{Universit\"{a}t Greifswald, Institut f\"{u}r Physik, 17487 Greifswald, Germany}

\author {M.D.~Seliverstov}
\affiliation{Petersburg Nuclear Physics Institute, NRC Kurchatov Institute, Gatchina 188300, Russia}

\author {P.~Spagnoletti}

\affiliation{School of Computing, Engineering, and Physical Sciences, University of the West of Scotland, Paisley PA1 2BE, United Kingdom}

\author {C.~Van~Beveren}
\affiliation{KU~Leuven, Instituut voor Kern- en Stralingsfysica, B-3001 Leuven, Belgium}

\author {M.~Veinhard}
\affiliation{CERN, CH-1211 Geneva 23, Switzerland}

\author{E.~Verstraelen}
\affiliation{KU~Leuven, Instituut voor Kern- en Stralingsfysica, B-3001 Leuven, Belgium}

\author{A.~Welker}
\affiliation{CERN, CH-1211 Geneva 23, Switzerland}
\affiliation{Institut f\"{u}r Kern- und Teilchenphysik, Technische Universit\"{a}t Dresden, Dresden 01069, Germany}

\author{K.~Wendt}
\affiliation{Institut f\"{u}r Physik, Johannes Gutenberg-Universit\"{a}t, D-55099 Mainz, Germany}

\author{R.N.~Wolf}
\thanks{Present address: ARC Centre of Excellence for Engineered Quantum Systems, School of Physics, The University of Sydney, NSW 2006, Australia.}
\affiliation{Max-Planck-Institut f\"{u}r Kernphysik, Saupfercheckweg 1, 69117 Heidelberg, Germany}
\affiliation{Universit\"{a}t Greifswald, Institut f\"{u}r Physik, 17487 Greifswald, Germany}

\author{A.~Zadvornaya}
\affiliation{KU~Leuven, Instituut voor Kern- en Stralingsfysica, B-3001 Leuven, Belgium}

\author{K. Zuber}
\affiliation{Institut f\"{u}r Kern- und Teilchenphysik, Technische Universit\"{a}t Dresden, Dresden 01069, Germany}

\date{\today}% It is always \today, today,
             %  but any date may be explicitly specified

\begin{abstract}
\noindent Combining laser spectroscopy in a Versatile Arc Discharge and Laser Ion Source, with Penning-trap mass spectrometry at the CERN-ISOLDE facility, this work reports on mean-square charge radii of neutron-rich mercury isotopes across the $N=126$ shell closure, the electromagnetic moments of $^{207}$Hg and more precise mass values of $^{206-208}$Hg.  The odd-even staggering (OES) of the mean square charge radii and the kink at $N=126$ are analyzed within the framework of covariant density functional theory (CDFT), with comparisons between different functionals to investigate the dependence of the results on the underlying single-particle structure. The observed features are defined predominantly in the particle-hole channel in CDFT, since both are present in the calculations without pairing. However, the magnitude of the kink is still affected by the occupation of the $\nu 1i_{11/2}$ and $\nu 2g_{9/2}$ orbitals with a dependence on the relative energies as well as pairing
\end{abstract}

\maketitle

\section{Introduction}
\label{Introduction}

The kink in the relative mean square charge radii ($\delta \langle r^2\rangle$) at the $N=126$ shell closure has long been considered as a benchmark for testing theoretical calculations. Traditionally the lead isotopic chain was employed~\cite{Tajima1993,Sharma1993,Reinhard1995,Fayans2000,Goddard2013}, but new experimental results in this region revealing the systemics of other isotopic chains~\cite{Cocolios2011,Farooq-Smith2016,Barzakh2018,Goodacre-2021}, mass measurements, and odd-even staggering (OES) in charge radii offer the opportunity to broaden this benchmark. The droplet model is unable to reproduce this kink because of the absence of single-particle degrees of freedom~\cite{Myers1983}. Early non-relativistic mean field approaches were also incapable of reproducing a kink at $^{208}$Pb~\cite{Thompson1982}, while conversely, relativistic mean field approaches were demonstrated to be successful in doing so~\cite{Sharma1993}. 

Two alternative modifications were suggested to correct this deficiency in non-relativistic models. The first relies on the modification of the spin-orbit interaction, either through a fitting procedure (see Refs.~\cite{Sharma1995,Reinhard1995}) or via the introduction of a density dependence (see Refs.~\cite{Nakada2015,Nakada2015a}). This leads to a reasonable reproduction of the experimental isotope shifts (see Refs.~\cite{Reinhard1995,Nakada2019}).
% but comes at the expense of a correct prediction of the masses~\cite{Fayans2000,Nakada-private}. 
The second approach (employing so-called Fayans functionals) introduces gradient terms into the pairing and surface terms of the functional~\cite{Fayans2000,Fayans1994,Reinhard2017}. This significantly improves the general description of experimental data, however, discrepancies are still apparent in the lead and tin isotopic chains~\cite{Gorges2019}. Moreover, pairing becomes	a dominant contributor to the kink and OES~\cite{Gorges2019}, in contradiction with the results obtained in relativistic Hartree-Bogoliubov (RHB) calculations with the DD-ME2 functional and non-relativistic Hartree-Fock-Bogoliubov (NR-HFB) calculations with the M3Py-P6a functional presented in Ref.~\cite{Goodacre-2021}.
	
This work is an in-depth follow-up to Ref.~\cite{Goodacre-2021}, which reported on the $\delta \langle r^2\rangle$ of mercury isotopes across $N$=126 and employed these results, together with existing lead data, to compare RHB and NR-HFB approaches. A new OES mechanism was additionally suggested, related to the staggering in the occupation of the different neutron orbitals in odd- and even-$A$ nuclei and facilitated by particle-vibration coupling (PVC) in odd-$A$ nuclei. Here we report on the magnetic-dipole and electric-quadrupole moments of $^{207}$Hg, new and improved mass measurements of $^{206-208}$Hg and a detailed theoretical study within the RHB framework to better understand the kinks and OES in lead and mercury isotopes. Multiple state-of-the-art covariant energy density functionals (CEDFs) are employed (NL3*~\cite{NL3*}, DD-PC1~\cite{DD-PC1}, DD-ME2~\cite{DD-ME2} and DD-ME$\delta$~\cite{DD-MEdelta}) to assess the dependence of the theoretical results on the underlying single-particle structure. The global performance of these functionals in describing ground state properties such as masses and charge radii of even-even nuclei has been tested in Refs.~\cite{Agbemava2014} and~\cite{AA.16}.

This article is arranged as follows. The experimental techniques are presented in Sec.~\ref{exper}. The radiogenic production of $^{207,208}$Hg in a molten target is discussed in Sec.~\ref{Radio-prod}. Experimental results on mean square charge radii, - dipole and electric-quadrupole moments and masses are summarized in Sec.~\ref{sec:results_discussion}. The discussion of experimental results is presented in Sec.~\ref{sec:DA}. Theoretical formalism and theoretical analysis of the kinks and OES in charge radii are presented in Sec.~\ref{theory-sec}, together with their dependence on the underlying single-particle structure and pairing and a comparison of experimental and calculated binding energies. Finally, we give a brief summary in Sec.~\ref{conclusion}.
%%%%%%%%%%%%%%%%%%%
\section{Experimental technique}
\label{exper}
%%%%%%%%%%%%%%%%%%%

\begin{figure}
\includegraphics[width=1\columnwidth]{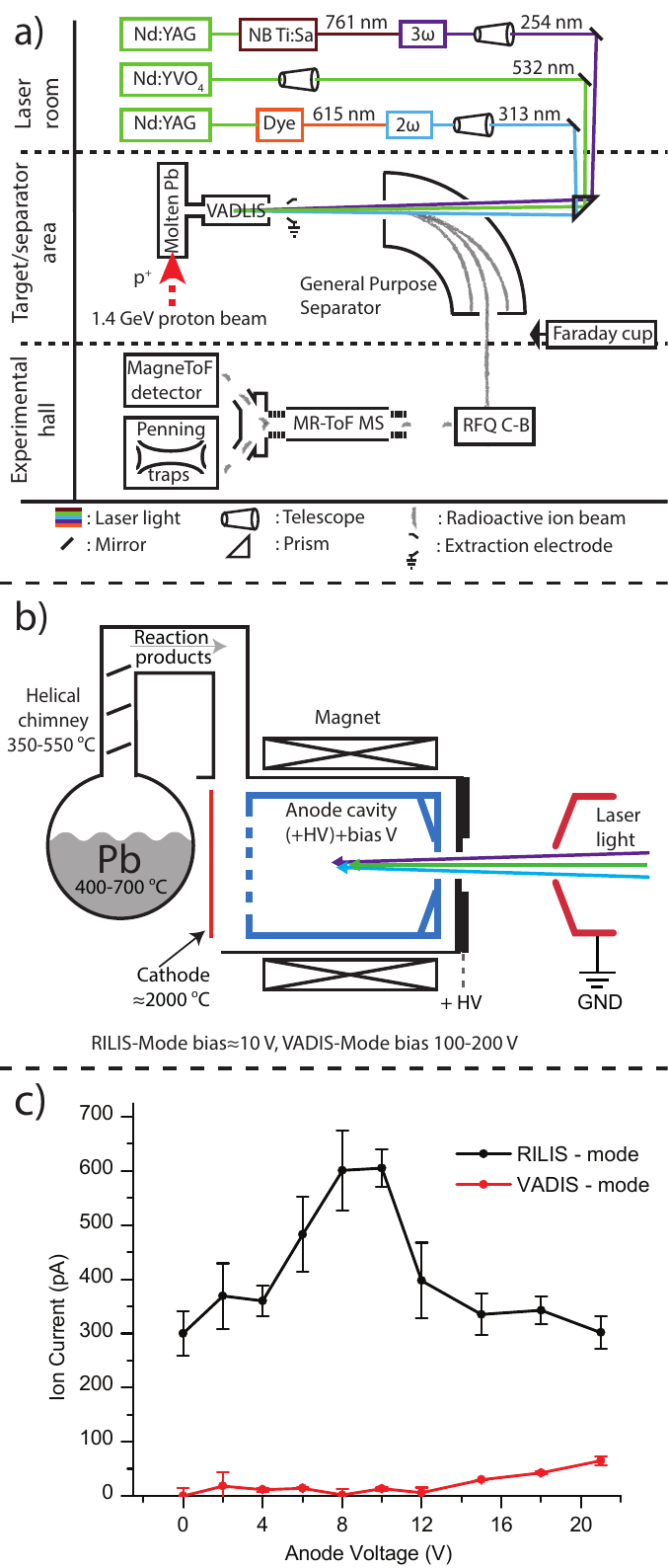}
	\caption [VADLIS principles and new operational modes.]{\footnotesize a) Overview of the experimental setup for the production and study of mercury isotopes. b) Schematic of a VADLIS coupled to a molten lead target via a temperature controlled chimney. c) The variation in the RILIS-mode and VADIS-mode ion currents on mass A=197 for differing anode bias voltages. See text for additional details and the definition of acronyms.} 
	\label{fig:VADLIS}
\end{figure}

\noindent The neutron-rich mercury isotopes were studied at the ISOLDE facility~\cite{Catherall2017} as part of a wider experimental campaign which investigated both ends of the isotopic chain~\cite{Goodacre-2021,Marsh2018,Sels2019}. The mercury nuclei were produced using the Isotope Separator On-Line (ISOL) method~\cite{Kofoed_Hansen1951,Blumenfeld2013} and studied via in-source resonance ionization spectroscopy~\cite{Alkhazov1992,Marsh2013} as depicted in Fig.~\ref{fig:VADLIS}(a).

A molten lead target (thickness 170 g/cm$^{2}$) was bombarded with 1.4-GeV protons, resulting in a cocktail of reaction products which effused via a temperature controlled chimney~\cite{Lettry1997a} into the anode volume of a Versatile Arc Discharge and Laser Ion Source (VADLIS)~\cite{DayGoodacre2016b}. The target and ion source were biased at 30~kV and laser light from the ISOLDE Resonance Ionization Laser Ion Source (RILIS)~\cite{Fedosseev2017} was directed into the anode volume for multi-step resonance ionization~\cite{Letokhov1987} of mercury isotopes. A $\{\lambda_{1}\left|\lambda_{2}\right|\lambda_{3}\} = \{254~$nm$\vert 313~$nm$\vert$532~nm$\}$ ionization scheme~\cite{DayGoodacre2017a} was applied, with the first (254~nm) resonant transition used to investigate the hyperfine structure (hfs) and isotope shifts in the 5d$^{10}$6s$^2$~$^1$S$_0$ $\rightarrow$ 5d$^{10}$6s6p~$^3$P$_1^{\circ}$ atomic transition.

The ions were accelerated by the electric field resulting from the grounded extraction electrode depicted in Fig.~\ref{fig:VADLIS}(a) and Fig.~\ref{fig:VADLIS}(b) to form a 30-keV radioactive ion beam (RIB). The ISOLDE General Purpose Separator~\cite{Catherall2017} was employed for mass separation before the RIB was directed to either a Faraday cup for direct ion current measurement or to the ISOLTRAP radio frequency quadrupole cooler-buncher (RFQ C-B)~\cite{Herfurth2001}. Downstream of the RFQ C-B, the RIB was injected into the Multi-Reflection Time-of-Flight Mass Spectrometer (MR-ToF~MS)~\cite{Wolf2013a} for either isobaric separation and subsequent detection~\cite{Wolf2012a} or for mass measurements, either by measuring the time-of-flight of the ions~\cite{Wienholtz2013} or by utilizing the downstream Penning traps~\cite{Wolf2013b}. 

The lead target-VADLIS combination was required to avoid the overwhelming isobaric francium contamination present on masses $A$=207, 208 when employing a standard UC$_{\text{x}}$ target with a hot cavity surface ion source for the laser light-atom interaction region~\cite{Lukic2006}. Alternative approaches have struggled to suppress certain isotopes of francium, and would additionally be expected to result in a factor of $\approx$20 reduction of the signal of interest~\cite{Fink2015,Fink2015a}.

A schematic of the VADLIS is presented in Fig.~\ref{fig:VADLIS}(b) together with the relative bias of the components. In the standard Versatile Arc Discharge Ion Source (VADIS)-mode of operation, atoms and molecules are ionized by electrons that are emitted from the $\approx$2000~$^{\circ}$C cathode and accelerated into the anode volume by a relative anode voltage of 100-200~V~\cite{Penescu2010}. The selective RILIS-mode of operation was employed for this experiment, where the anode voltage is optimized for laser-ion extraction while maintaining it below what is required for significant electron impact ionization~\cite{DayGoodacre2016b}. Figure~\ref{fig:VADLIS}(c) presents the on-line optimization of the RILIS-mode with radiogenically produced $^{197}$Hg. The RILIS-mode and VADIS-mode related signals were separated by blocking and unblocking the laser light exciting the 254-nm transition. A clear maximum is visible with a near-negligible background with the anode voltage set to $\approx$8~V. The alternative, applying the RILIS lasers with a 100-200~V anode bias (termed RILIS+VADIS-mode) would have resulted in significant isobaric contamination and a reduced signal-to-noise ratio as a result of the competing ionization processes.

\begin{figure}%
	\includegraphics[width=1\linewidth]{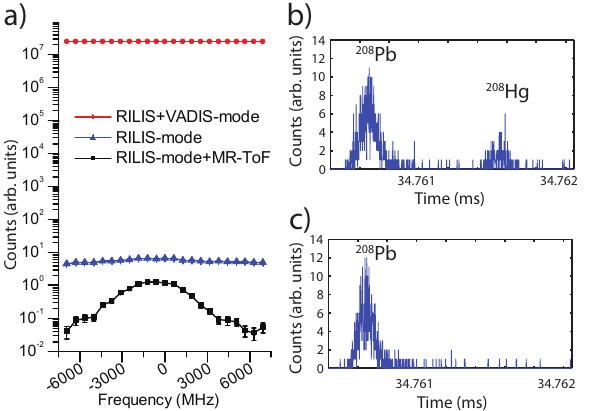}%
	\caption[MR-ToF~MS time-of-flight spectra on mass A=208]{\footnotesize a) Ion rate on mass $A$=208 measured by the MR-ToF MS detector for a given frequency tripled laser frequency. The RILIS+VADIS-mode is dominated by VADIS ionized $^{208}$Pb, the count rate was estimated based on Faraday cup measurements. Further discussion in the text. b) and  c) time-of-flight spectra on mass $A$=208, measured downstream of the MR-ToF~MS with the VADLIS in RILIS-mode and the lasers on resonance and off-resonance, respectively. The y-axis scales of b) and c) are identical.
	}%
	\label{fig:ToF}%
\end{figure}

The benefits of combining the RILIS-mode of operation with the ISOLTRAP MR-ToF MS are highlighted in Fig.~\ref{fig:ToF}(a). Operating in RILIS-mode reduced the isobaric $^{208}$Pb background by seven orders of magnitude compared with RILIS+VADIS-mode. This enabled the MR-ToF~MS to be employed for selective detection and for determining the mass of $^{208}$Hg. Time-of-flight spectra recorded on and off resonance with the MR-ToF~MS are shown in Fig.~\ref{fig:ToF}(b) and \ref{fig:ToF}(c), respectively. By applying time gates in the ToF spectra, it was possible to separate the $^{208}$Hg signal from the remaining isobaric contamination.

\section{Radiogenic production of $^{207,208}\text{Hg}$ in a molten lead target}\label{Radio-prod}
Considering the natural lead ($^{206-208}$Pb) target material used for this experiment, the creation of mercury isotopes with $N\leq 126$ is comparatively well understood as the result of spallation reactions induced by the incident 1.4~GeV proton beam. However, when going beyond $N=126$ ($^{206}$Hg) the production mechanism changes, and a range of other processes may become relevant~\cite{Tall2007,Rodriguez-Sanchez2020} including secondary reactions induced by the light and energetic products of the primary spallation reactions. The production of $^{207}$Hg via $^{208}$Pb($n,2p$)$^{207}$Hg is a good example of such a process, and was first reported at an ISOL facility in Ref.~\cite{Jonson1981}. 
The mechanism for producing $^{208}$Hg is significantly more exotic, as evidenced by a factor of 2400 decrease in the measured ion rate between $^{207}$Hg and $^{208}$Hg. There are a number of potential production channels including $^{208}$Pb(t,3$p$)$^{208}$Hg, $^{208}$Pb($\alpha$,4$p$)$^{208}$Hg, or reactions with radiogenically produced $^{209}$Pb (t$_{1/2}\approx$~3~h) or $^{210}$Pb (t$_{1/2}\approx$~22~y) which build up within the target during the experiment.

\begin{figure}%
	\includegraphics[width=1\linewidth]{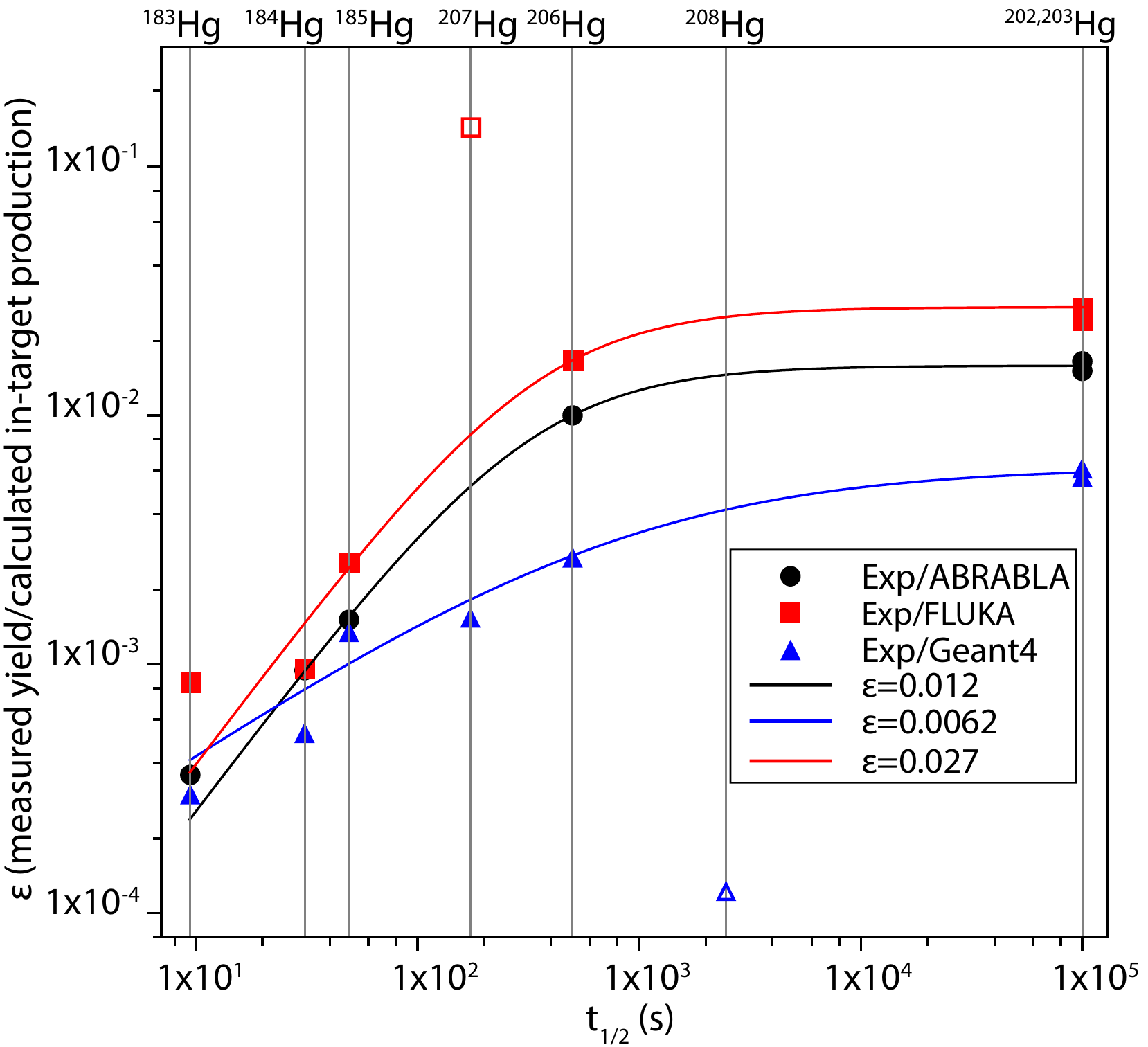}%
	\caption[Ionization and release efficiency ($\varepsilon$) as a function of the half-life of mercury isotopes from a molten lead target determined using results from ABRABLA, FLUKA and Geant4 simulations. See text for details.
]{\footnotesize Ionization and release efficiency ($\varepsilon$) as a function of the half-life of mercury isotopes from a molten lead target. The results from ABRABLA, FLUKA and Geant4 simulations are compared. See text for details.
	}%
    \label{fig:Prod}%
\end{figure}

The in-target production of mercury isotopes was calculated via ABRABLA~\cite{Gaimard1991,Junghans1998}, FLUKA~\cite{Ferrari2005,Bohlen2014} and Geant4~\cite{Agostinelli2003,Allison2016,Garcia2017} simulations. The results are assessed by considering the isotope specific extraction and ionization efficiencies~($\varepsilon$), determined by dividing the measured yield by the calculated in-target production. Figure~\ref{fig:Prod} presents the relationship between $\varepsilon$ and half-life for the mercury isotopes measured with the MR-ToF during this experimental campaign. For $^{202}$Hg (stable) and $^{203}$Hg (t$_{1/2}\approx$47 days) the half-lives are set at 1$\times 10^5$~s to facilitate their inclusion. The data is fitted using Eq.~(4) from Ref.~\cite{Lukic2006}, with the hollow data points omitted from the fits to enable them to converge. As expected, $\varepsilon$ generally increases with increasing half-life, and stabilizes at a point where the half-life is sufficiently long compared to the release time. 

All of the results broadly agree with an extraction and ionization efficiency of approximately 1\% for sufficiently long-lived isotopes. ABRABLA~\cite{Gaimard1991} is not capable of reproducing the secondary reactions required for the production of $^{207,208}$Hg, however, as it is commonly used for calculating in-target production, it is useful for benchmarking the other codes for this application. FLUKA~\cite{Bohlen2014} was found to reproduce some $^{207}$Hg production, though based on Fig.~\ref{fig:Prod} the rate appears to be underestimated. The Geant4 (geant4$-$10$-$07) simulations employing the Liege (QGSP$\_$INCLXX$\_$HP) model~\cite{Allison2016} combined with the native de-excitation code were the most successful in reproducing both $^{207}$Hg and $^{208}$Hg production, though with an apparent overestimation of the latter. Discrepancies with the Geant4 results may be a consequence of the necessity to scale from the simulation of a reduced density target, which was required to enable a feasible simulation time. 

While $^{208}$Hg production was not present in the FLUKA results, the simulations presented the possibility to investigate the $^{208}$Pb(t,3$p$)$^{208}$Hg channel. The results for tritium production (using FLUKA2021.0) were convoluted with cross-section data from TENDL17~\cite{Koning2012} over a 40-200~MeV interval. This resulted in an in-target production rate of 110~atoms/\textmu C for $^{208}$Hg, significantly below the estimated rate of $\approx$56,000~atoms/\textmu C calculated considering a 1$\%$ extraction and ionization efficiency. This suggests that $^{208}$Pb(t,3$p$)$^{208}$Hg reactions only contribute to a fraction of the observed $^{208}$Hg yields. Based on this, we tentatively conclude that the observed $^{208}$Hg production is the result of a combination of multiple reaction channels.

\section{Experimental Results}\label{sec:results_discussion}
\subsection{Laser spectroscopy of mercury isotopes across $N$=126} 
\begin{figure*}
	\includegraphics[width=1\linewidth]{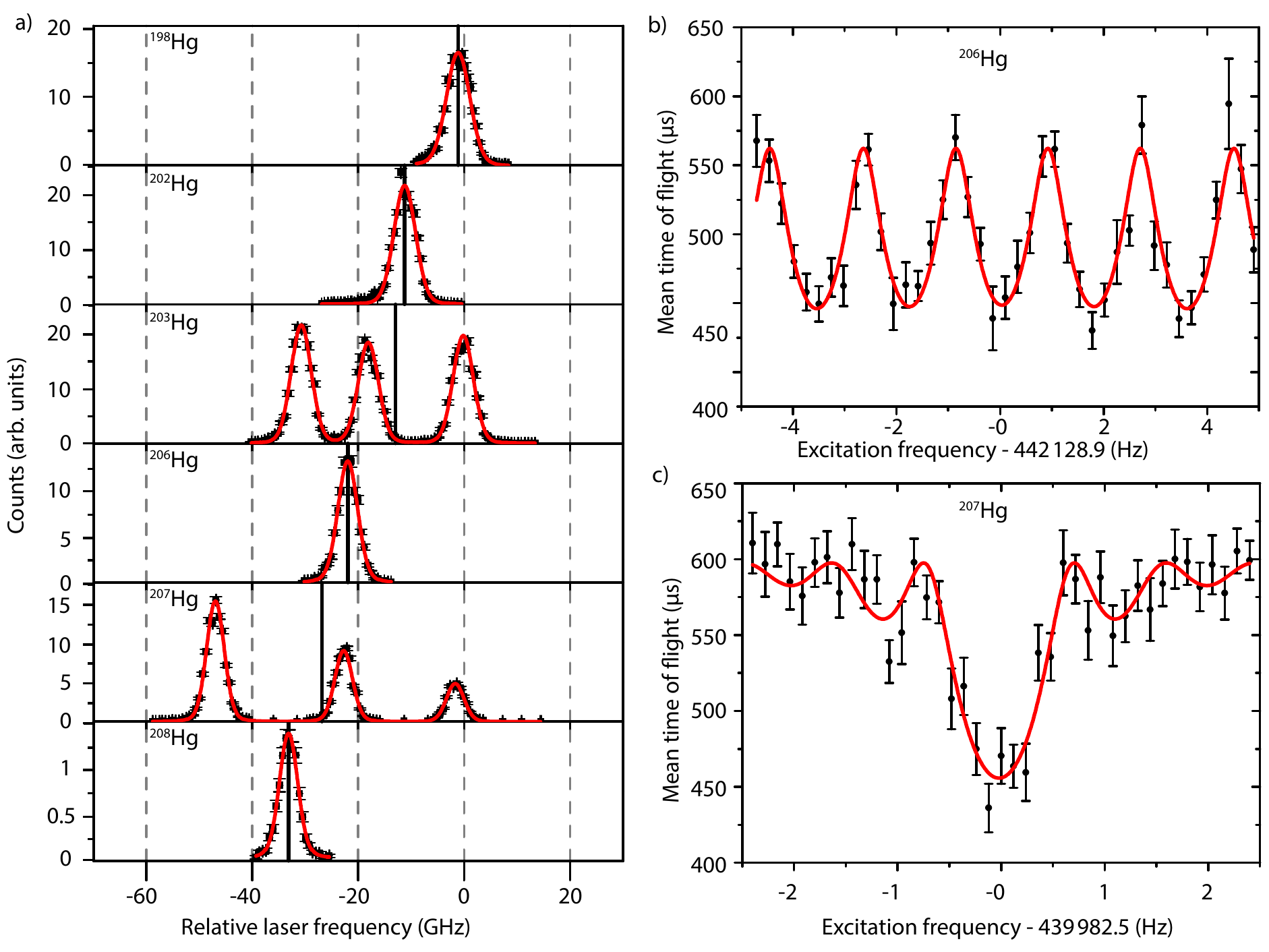}
	\caption [VADLIS principles and new operational modes.]{\footnotesize (a) Hfs spectra of the measured isotopes, the (substate weighted) centroids are indicated with solid black lines. The y-axis represents the number of ``counts per shot" from the RFQ C-B. (b) and c) Sample plots of the measured time-of-flight ion-cyclotron resonances. (b) Measurement of $^{206}$Hg using a Ramsey-type excitation scheme with an excitation time of $2\times\SI{60}{\milli\second}$ separated by $\SI{480}{\milli\second}$. 
	(c) Measurement of $^{207}$Hg using a single-pulse excitation scheme with an excitation time of $\SI{1.2}{s}$.} 
	\label{fig:laser_scans}
\end{figure*}

Mean square charge radii, magnetic-dipole and electric-quadrupole moments were studied via the measurement of isotope shifts and hfs in the 254-nm transition. Sample spectra are presented in Fig.~\ref{fig:laser_scans}(a), with the (substate weighted) centroids indicated with solid black lines.

Reference scans of $^{198}$Hg were taken periodically to monitor the stability of the experimental setup, with a 10~h interval. Multiple measurements of the hfs of each isotope were taken (2$\times ^{202,203}$Hg, 3$\times ^{206,207}$Hg, 5$\times ^{208}$Hg) and the fitting of the resulting spectra was cross-checked using multiple software packages: Origin 2016 with a chi-squared minimization~\cite{OriginLab} performed with a Levenberg-Marquardt algorithm~\cite{Levenberg1944,Marquardt1963}, the SATLAS open source Python package~\cite{Gins2018} and a similar program written in ROOT~\cite{Brun1997}. The results are presented in Table~\ref{tab:Exp_results}, together with literature data ($^{202,203,206}$Hg) for comparison. Relative mean square charge radii, electric dipole and magnetic quadrupole moments were extracted from the spectra by applying standard methods, these are summarized in Appendix~\ref{ch:app:NO_HFS}.

\begin{table*}
	
	\centering
	\caption[Final results]{\footnotesize Extracted isotope shifts in the 254-nm line with respect to $^{198}$Hg, hyperfine $a$ and $b$ factors of the 5d$^{10}$6s6p~$^{3}$P$_{1}^{\circ}$ state and literature values recalculated from the compilations of~\cite{Ulm1986} and~\cite{Bonn1976}. The spin assignment of $^{207}$Hg is discussed in the text. Statistical uncertainties are listed in parentheses and the systematic uncertainties related to $F_{254}$ and $M$ are listed in curly brackets.} 
	\vspace{+5mm}
	
	\begin{tabular}{ c c c c c c c c c} \hline
		
		Isotope&$I^{\pi}$&$\delta v^{A, 198}$ & $a$ & $b$ &$\delta \langle r^2\rangle ^{A, 198}$& $\mu$ &$Q_s$&Ref. \\
	
		A&&(MHz)&(MHz)&(MHz)&(fm$^2$)&($\mu _N$)&($b$)&\\\hline
		202&0& -10 100(180) &  & &0.197(3)\{14\}	&&&This work \& \cite{Goodacre-2021} \\
		 && -10 102.4(4.2) & & &0.1973(2)\{152\}&&&\cite{Ulm1986}\\ 
		203&5/2$^{-}$& -11 870(200) & 5070(90) &-20(250)&0.232(5)\{17\}&0.843(15)&0.03(35)& This work \& \cite{Goodacre-2021}\\
		 && -11 750(180) & 4991.33(4) & -249.2(3)&	0.2296(35)	\{180\} &0.8300(7) &0.40(4)&\cite{Ulm1986,Bonn1976}\\ 
		 
		206&0& -20 930(160) & & &0.409(3)\{30\} & & &This work \& \cite{Goodacre-2021}\\
	 	 &&  -20 420(80) & & &0.3987(16)\{308\}& & &\cite{Ulm1986}\\
		207&(9/2$^{+}$)&-25 790(190)&-4500(60)&530(250)&0.503(4)\{38\}&-1.373(20)&-0.73(37)&This work \& \cite{Goodacre-2021}\\
		208&0&-32 030(160)&&&0.625(3)\{47\}&&&This work \& \cite{Goodacre-2021}\\  
		
	\end{tabular}
	
	\label{tab:Exp_results}
\end{table*}

The nuclear spin of $^{207}$Hg could not be determined unambiguously because the spectroscopic transition is between atomic states with electronic spins $J$=0 and $J$=1. $I^{\pi}=9/2^{+}$ was assumed for the analysis of the $^{207}$Hg measurements based on Refs.~\cite{Jonson1981,Tang2020}. The extracted isotope shifts for $^{202,203}$Hg are in good agreement with literature. The same is true for the neutron deficient isotopes that were re-measured during this experimental campaign~\cite{Marsh2018,Sels2019}. The 500-MHz discrepancy between the $\delta \nu^{206, 198}$ value of~\cite{Dabkiewicz} and this work is discussed in~\cite{DayGoodacre2016}. The general agreement of the extracted $\delta v^{A, 198}$ and hyperfine $a$ and $b$ factors with the previously published literature values further validates the method of in-source resonance ionization spectroscopy with a VADLIS ion source.

\subsection{Mass spectrometry of $^{\mathit{206-208}}$Hg}\label{ch:mass_spectrometry}
\begin{figure}%
	\includegraphics[width=1.0\linewidth]{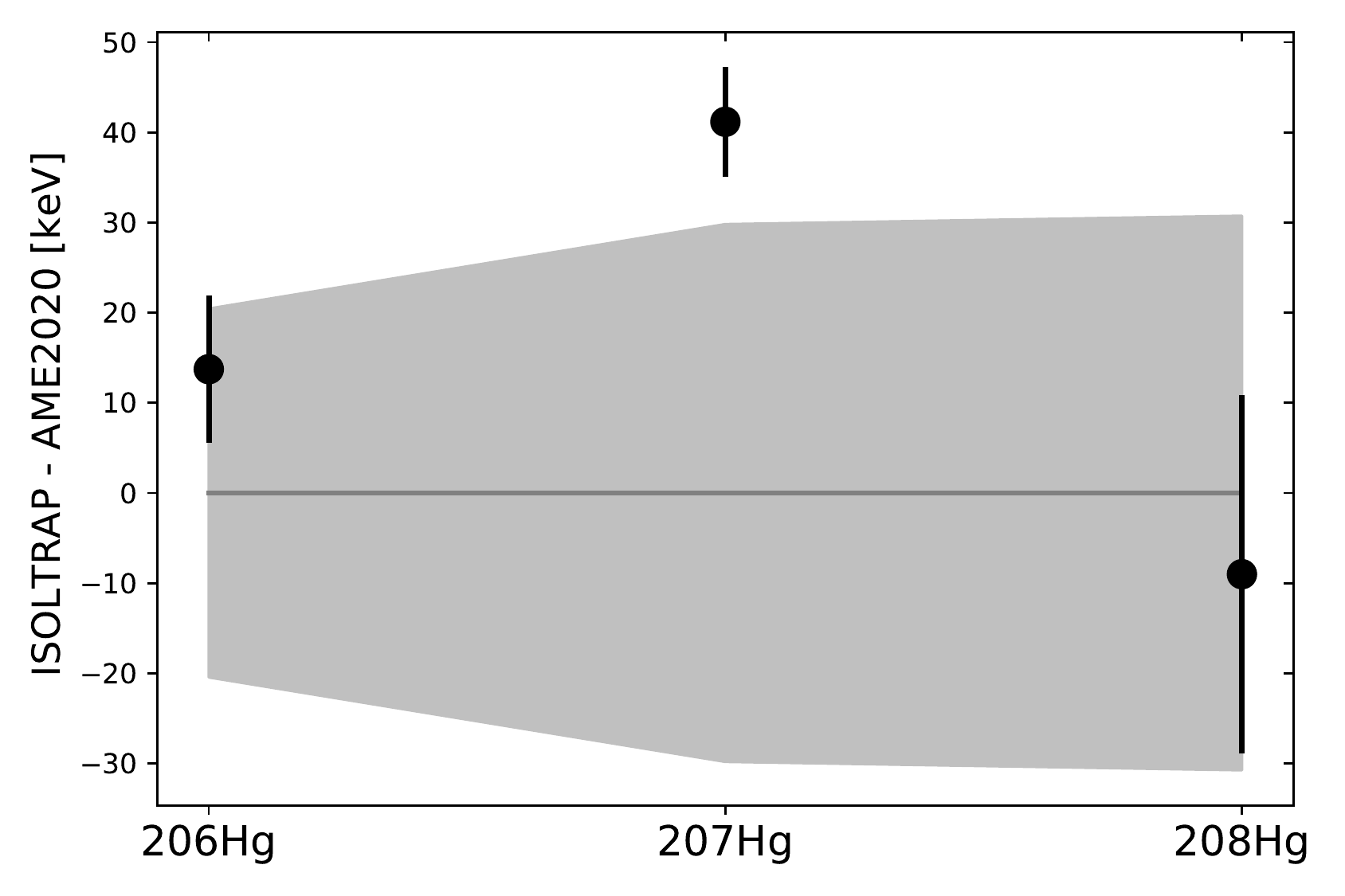}%
	\caption[Mass measurement comparison]{\footnotesize Difference between the ISOLTRAP mass measurements of this work and the corresponding AME20 values (black), including the AME20 error band (grey).}
	\label{fig:Mass_comparison}%
\end{figure}

The masses of $^{206-208}$Hg were measured, employing different techniques with respect to earlier experiments that are referenced in the AME2020 \cite{AME2020}. A short outline of the time-of-flight methods that were used in our experiment can be found in Appendix~\ref{ch:app:mass_measurements}. A summary of the measured values is presented in Table \ref{tab:Exp_results_masses} and a comparison with the AME2020 is shown in Figure \ref{fig:Mass_comparison}.

The atomic mass of $^{206}$Hg was previously deduced from $\alpha$-decay measurements~\cite{Kauranen1962}. Our measurements of $^{206}$Hg, using the Time-of-Flight Ion-Cyclotron-Resonance (ToF-ICR) technique, represent the first direct determination of the mass of this nucleus. For the ToF-ICR measurements in a Penning trap, two excitation schemes were employed: a single pulse excitation of $\SI{400}{\milli\second}$, as well as a Ramsey-type scheme~\cite{George2007}, presented in Figure~\ref{fig:laser_scans}b, where two excitation pulses of $\SI{60}{\milli\second}$ were applied, separated by a waiting time of $\SI{480}{\milli\second}$.

The masses of $^{207}$Hg and $^{208}$Hg have been determined previously by storage-ring measurements at GSI using Schottky mass spectrometry~\cite{Chen2008, Chen2009,Chen2012}. In the present experiment, ToF-ICR measurements were performed for $^{207}$Hg with a single-pulse excitation of $\SI{1.2}{\second}$ (see Figure \ref{fig:laser_scans}c). For $^{208}$Hg, the hyperfine structure was measured in five different laser scans, using ISOLTRAP's MR-ToF MS as mass separator and ion counter. Mass data were extracted from the scans by summing individual binned data per scan step into a single histogram, resulting in one histogram per scan. The ToF-distributions corresponding to singly charged ions of the isotope of interest, $^{208}$Hg$^{+}$, and to an on-line reference ion, $^{208}$Pb$^{+}$, were aligned as outlined in \cite{Fischer2018} and fitted by employing an unbinned maximum-likelihood estimation where the fit function was constructed as an Exponential-Gaussian-Hybrid (EGH) to account for the tails of the ToF-distributions towards longer flight times. The mass was extracted by calculating the average $C_{\text{ToF}}$ of the five scans, using the $^{208}$Pb$^{+}$ present in the RIB and $^{133}$Cs$^{+}$ from an off-line ion source as reference ions. The results we report are in general agreement with the literature and improve upon the precision. 

\begin{table*}

	\centering
	\caption[Mass results]{\footnotesize Mass-measurement results for the mercury isotopes, given either as the ratio $R$ of cyclotron frequencies from ToF-ICR or as the C$_\text{ToF}$ from the MR-ToF MS. The computed mass excess values M$_{\text{exc}}$ from this work are compared to the literature values found in \cite{AME2020}. Additionally, half-lives T$_{1/2}$ from \cite{AME2020} are given as well as the reference ions that were used to extract the mass values. The mass excesses are given in terms of energy divided by the square of the speed of light $\si{\clight}$.}
	%\vspace{+5mm}
	
	{\color{black}
	\begin{tabular}{c c c c c c c c c} \hline
		Isotope & T$_{1/2}$ & Ref. ions & C$_{\text{ToF}}$ & R & M$_{\text{exc, ISOLTRAP}}$ &M$_{\text{exc, AME20}}$& $|\Delta_\text{TRAP-AME20}|$ \\
		 & ($\si{\minute}$) &  &  & &($\si{\kilo\electronvolt/\clight\squared}$) &($\si{\kilo\electronvolt/\clight\squared}$)&($\si{\kilo\electronvolt/\clight\squared}$)&\\\hline
		$^{206}$Hg & $\SI{8.32\pm 0.07}{}$ & $^{133}$Cs$^+$ & / & $\SI{1.5498072282\pm0.0000000661}{}$  & $\SI{-20932.1\pm8.2}{}$ & $\SI{-20946\pm 20}{}$ & $\SI{13\pm 22}{}$  \\
		$^{207}$Hg & $\SI{2.9\pm 0.2}{}$ & $^{133}$Cs$^+$ & / & $\SI{1.5573676396\pm0.0000000489}{}$ & $\SI{-16446.2\pm 6.1}{}$ & $\SI{-16487\pm 30}{}$ & $\SI{41\pm 31}{}$ \\
		$^{208}$Hg & $\SI{42\pm 5}{}$ & $^{208}$Pb$^+$, $^{133}$Cs$^+$ & $\SI{0.500108969\pm 0.000000255}{}$ & / & $\SI{-13279\pm20}{}$ & $\SI{-13270\pm 30}{}$ & $\SI{9\pm36}{}$\\
	\end{tabular}}
	\label{tab:Exp_results_masses}

\end{table*}

\section{Discussion}\label{sec:DA}

\subsection{Magnetic moment of $^{\mathit{{207}}}$Hg}
\begin{figure}%
	\includegraphics[width=1\linewidth]{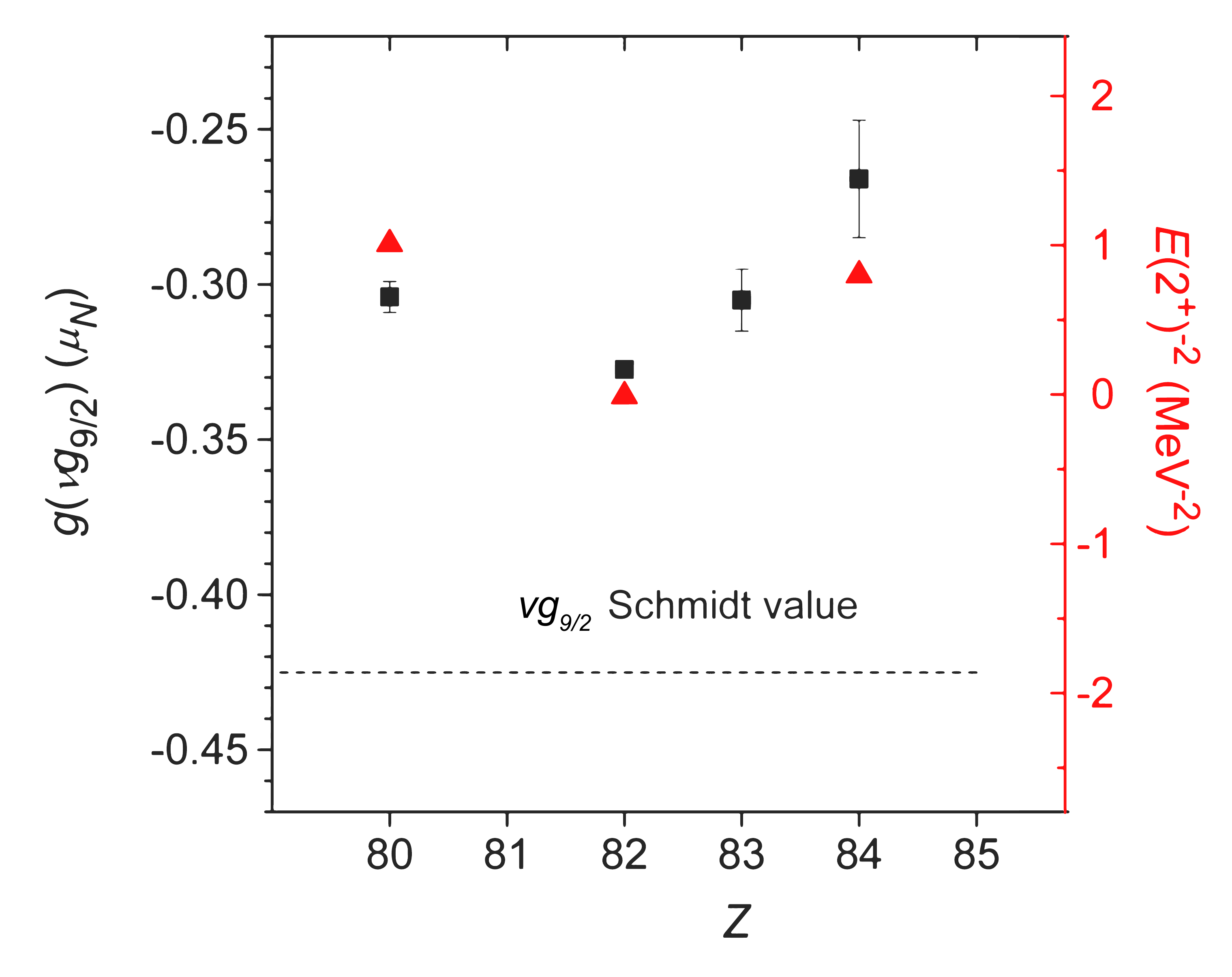}%
	\caption[N=127 $g$ factors]{\footnotesize Squares: experimental $g$-factors for $N=127$ isotones ($\nu g_{9/2}$~\cite{Baba1972,Pearson2000,Anselment1986,Seliverstov2014} and this work). The $g$-factor of $^{210}$Bi is deduced from measurements of isomer states, see text for details. Dashed line shows the Schmidt value for $\nu g_{9/2}$ shell. Triangles: $E$(2$^{+}$)$^{-2}$ values of the corresponding $N=126$ even-even nuclei (see the corresponding scale on the right-hand side of the figure) experimental data from~\cite{Pritychenko2016}.}
	\label{fig:N127g}%
\end{figure}

A value of $g$($^{207}$Hg)=$-$0.305(6) was deduced based on the results presented in Table~\ref{tab:Exp_results}. The $g$-factors of the $\nu g_{9/2}$ isotones $^{209}$Pb, $^{211}$Po are plotted in Fig.~\ref{fig:N127g} together with the Schmidt value, the energies of the first excited $2^+$ states ($E(2^+)^{-2}$) of the $N$=126 cores and $g(^{210}$Bi) calculated from the measured magnetic moments of the 5$^{–}$, 7$^{–}$ and 9$^{–}$ isomeric states~\cite{Baba1972,Pearson2000} using the additivity relation and assuming a pure $[\pi h_{9/2} \otimes \nu g_{9/2}]$ configuration for these states. The inverse square of $E$(2$^{+}$) is of particular relevance due to its approximate proportionality to the second-order perturbation theory correction~\cite{Heyde1992,Arima1986}.

The experimentally determined $g$-factors differ significantly from the Schmidt value $g_{Schmidt}$=$-$0.425 and there is a noticeable $Z$-dependence. The deviation of the $g$-factor of the magic $^{209}$Pb isotope was explained in non-relativistic~\cite{Arima1987,Arima1972} and relativistic~\cite{Li2013} approaches by taking into account corrections for the meson exchange current and first and second order core polarization (CP). The configuration admixture contributing to the magnetic dipole moment in the first order of perturbation theory corresponds to a particle-hole excitation from an orbit $j=l+1/2$ to its spin-orbit partner $j=l–1/2$ (CP1 correction~\cite{Arima1954}). In the second order of perturbation theory, the most important magnetic moment correction stems from the odd-particle coupling with the lowest 2$^{+}$ excitation of the core (CP2 correction~\cite{Heyde1992,Arima1986}). In the vicinity of the doubly magic $^{208}$Pb, the most important $Z$-dependent CP1 correction corresponds to the proton ($h^{-1}_{11/2}h_{9/2}$) particle-hole excitation. A corresponding increase in the occupancy of the $\pi h_{9/2}$ orbital with the increase of $Z$, reduces the probability of proton ($h^{-1}_{11/2}h_{9/2}$) core excitations, thereby decreasing the magnitude of the CP1 correction. The opposite is apparent in Fig.~\ref{fig:N127g}, where the deviation from the Schmidt value increases between $^{209}_{82}$Pb and $^{211}_{84}$Po, thus CP1 corrections do not appear to be the dominant driver for the $g$-factor $Z$-dependence. Additionally, meson exchange corrections have been shown to have a weak $A$ dependence in the vicinity of $^{208}$Pb~\cite{Arima1986,Towner1977}. It could then be suggested that CP2 corrections are primarily responsible for the $Z$-dependence of the discrepancy with the Schmidt value.

The same mechanism (particle-quadrupole-vibration coupling) was used to explain the magnetic moment evolution in the vicinity of the magic numbers~\cite{Arima1986, Heyde1992}. This explanation is supported by the apparent correspondence of the energies of the first excited $2^+$ states of the $N$=126 cores and the $g$-factors of the $N$=127 isotones in Fig.~\ref{fig:N127g}. Considering that $E$($2^+,^{208}$Pb)$\approx$4.1~MeV, $E$($2^+,^{206}$Hg)$\approx$1.1~MeV and $E$($2^+,^{210}$Po)$\approx$1.2~MeV~\cite{Pritychenko2016}, the admixture of the (2$^+$,$\nu i_{13/2}$)$_{9/2+}$ should be larger for $^{207}$Hg and $^{211}$Po than for $^{209}$Pb, resulting in an increased CP2 correction. It is worth noting that this interpretation indicates the importance of particle-vibration coupling in the description of the ground-state properties of the odd-$A$ nuclei in the vicinity of shell closure. This same mechanism proves to be decisive for explanation of the charge radii behavior in this region of the chart of the nuclides (see Sec.~\ref{theory-sec}).

\subsection{Quadrupole moment of~$^{\mathit{207}}$Hg}

Quadrupole moments near the closed proton and neutron shells have a predominantly single-particle nature and are usually well described by the shell-model formula:
\begin{equation}
Q_{}=-e_{eff} \frac{2j-1}{2j+1}\left\langle r^2 \right\rangle _j
\label{eq:NEM}
\end{equation}
\noindent where $j$ is the spin of the odd particle, $\langle r^2\rangle _j$ is its mean square radius and $e_{eff}$ is an effective charge. Taking $\langle r^2\rangle _j$ from Ref.~\cite{Sagawa1988}, one obtains: $e_{eff}(^{207}$Hg$; \nu 2g_{9/2})$=2.4(13) $e$. This neutron effective charge is noticeably larger than the value of the ``universal" neutron $e_{eff}=0.95~e$ which describes fairly well the measured quadrupole moments for all closed-shell $\pm 1$ nuclear states in the vicinity of $^{208}$Pb~\cite{Neyens2003}. This points to the rapid increase of a quadrupole core polarization when moving away from the magic $Z$=82, similar to that observed for the proton $e_{eff}$ when moving away from the magic $N$=126~\cite{Neyens2003}. 

%%%%%%%%%%%%%%%%%%%%%%%%%%
\section{Theoretical interpretation}
\label{theory-sec}
%%%%%%%%%%%%%%%%%%%%%%%%%%

%%%%%%%%%%%%%%%%%%%%%%%%%%%%%%%%%
\subsection{Theoretical framework and the details of the calculations}
%%%%%%%%%%%%%%%%%%%%%%%%%%%%%%%%%

A theoretical interpretation of experimental data is performed within the framework of covariant density functional theory \cite{Vretenar2005} using the RHB computer code for spherical nuclei first employed in Ref.~\cite{Goodacre-2021}. This code enables the blocking of selected single-particle orbitals and  allows for fully self-consistent calculations of the ground and excited states in even-even and odd-$A$ spherical nuclei. In the pairing channel, the RHB code employs a  separable version of the Gogny pairing ~\cite{Tian2009} with the pairing strength defined in Ref.~\cite{Agbemava2014}.

Several restrictions/constraints are employed in the present paper. First, we consider only spherical nuclei (i.e. for which $\left<\beta_2^2\right>^{1/2} <0.1$ where $\left<\beta_2^2\right>^{1/2}$ is the  mean-square deformation deduced from experimental $\delta \left<r^2\right>$ using the droplet model (see Refs.~\cite{Otten1989,Berdichevsky1985}).) This restriction corresponds to $N\geq 116$ and $N\geq 121$ for lead and mercury isotopes, respectively,  
and it was already used in our earlier paper~\cite{Goodacre-2021}. 

\begin{figure*}
\centering
\includegraphics[width=14.0cm]{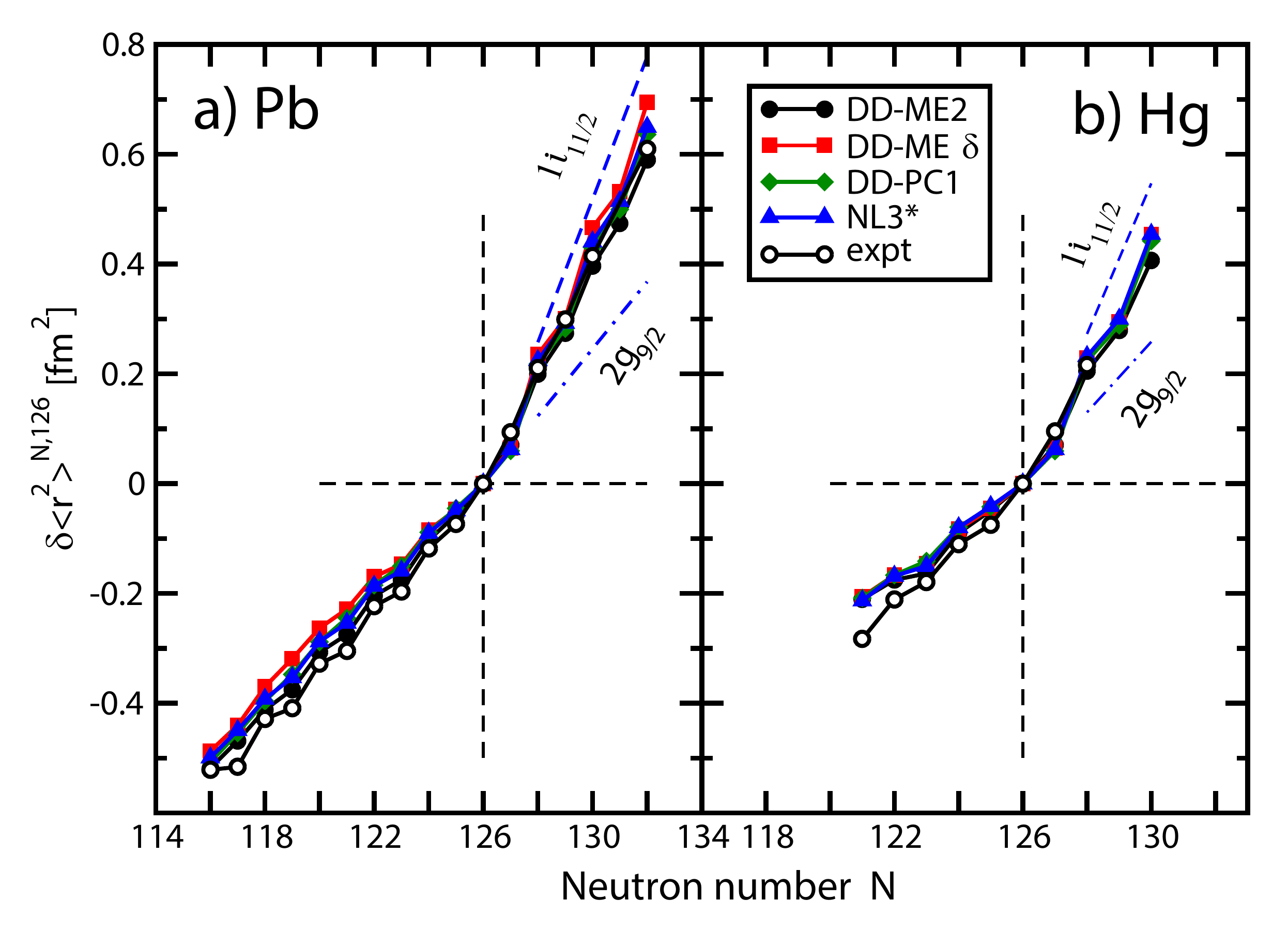}
	\caption{\footnotesize The $\delta \left< r^2 \right>^{N,126}$ values of the Pb and Hg isotopes 
		obtained in the RHB calculations with indicated CEDFs. 
		The experimental data are taken from Refs.~\cite{Goodacre-2021,Ulm1986,Anselment1986}. The 
		$\delta \left < r^2 \right>^{N,126}$ values, obtained in the calculations with CEDF NL3* without 
		pairing under the condition that either only $\nu 1i_{11/2}$  or only $\nu 2g_{9/2}$  orbitals are 
		occupied in the $N>126$ nuclei,  are shown by blue dashed and dash-dotted lines, respectively. 
	}
	\label{fig:Pb-Hg-diff-charge-radii}%
\end{figure*}

Second, two different procedures labeled as “LES” and “EGS” are used for the blocking in odd-$A$ nuclei, and the results of the respective calculations are labeled by these abbreviations. In the LES (lowest in energy solution) procedure, the lowest in energy configuration is used, which is similar to all earlier calculations of OES in non-relativistic DFTs~\cite{Fayans2000,Reinhard2017}. In the EGS (experimental ground state) procedure, the configuration with the spin and parity of the blocked state corresponding to those of the experimental ground state is employed, although it is not necessarily the lowest in energy. The need for this procedure is due to the following considerations. First, the nodal structure of the wavefunctions  and neutron radius of the single-neutron orbital depend on its quantum numbers such as total and orbital angular momenta	(see Refs.~\cite{Reinhard1995,Goddard2013,UAR.21}). Thus, the occupation of different neutron single-particle states impacts the resulting charge radii (see Refs.~\cite{Reinhard1995,Goddard2013} and the discussion of Fig.~\ref{fig:Pb-Hg-diff-charge-radii}(a) below). Second, the structure of the experimental ground states in odd-$A$ nuclei is reproduced globally only in approximately 40\% of the nuclei in the non-relativistic and relativistic DFTs~\cite{BQM.07,AS.11}\footnote{The  inclusion of  particle-vibrational coupling increases the accuracy of the description of the single-particle configurations in odd-$A$ nuclei (see Refs.~\cite{Litvinova2011,Afanasjev2015}).}. 

If there is a mismatch in the structure of experimental and calculated ground states the impact of the blocked orbital on the physical observable of interest (charge radius, deformations, binding etc.) in odd-$A$ is expected to deviate from the value observed in experiment. The consequences of this mismatch exist for the deformations of one-quasiparticle states (see Ref.~\cite{AS.11}), odd-even staggerings in charge radii (see Ref.~\cite{Goodacre-2021}) and binding energies (see Ref.~\cite{TA.21}). Note that the latter is used to define pairing indicators. In such a situation it is safer to use the blocked solution with the structure corresponding to experimental ground state (even if it leads to an excited solution) for the description of charge radii since moderate shift in the energy of single-particle state  has negligible effect on its neutron single-particle rms radius $\left<r^{2}\right>_{sp}$.

%%%%%%%%%%%%%%%%%%%%%%%%%%
\subsection{Charge radii and related indicators}
%%%%%%%%%%%%%%%%%%%%%%%%%%

The charge radii were calculated from the corresponding point proton 
radii as
\begin{equation}
r_{ch} = \sqrt{\left< r^2 \right> + 0.64}\,\,\,\, {\rm fm}
\end{equation}
where $\left< r^2 \right>$ stands for mean square radius of proton density 
distribution and the factor 0.64 accounts for the finite-size  of the proton.

%%%%%%%%%%%%%%%%%%%%%%%%%%%%%%%%%%%%%%%%%%%%%%
%%%%%%%%%%%%%%%%%%%%%%%%%%%%%%%%%%%%%%%%%%%%%%%

Three indicators are commonly used  to facilitate  the quantitative comparison of the experimental results with those from theoretical calculations. 
%One of them 
The first is differential mean-square  charge radius\footnote{This quantity 
	is frequently written as a function of  mass number $A$. However, we prefer to define 
	it as a function of neutron  number $N$ since this allows to see the behavior of the 
	$\delta \left < r^2 \right>^{N,N'}$ curves at neutron shell  closures for 
	different isotopic chains.} 
\begin{eqnarray}
\delta \left < r^2 \right>^{N,N'} = \left< r^2 \right>(N) 
- \left < r^2 \right>(N') = 
r^2_{ch}(N) - r^2_{ch}(N')
\end{eqnarray}
where $N'$ is the neutron number of the reference nucleus.  

The second one is the $\xi_{even}$ indicator 
\begin{eqnarray}
\xi_{even}= \frac{\delta\langle r^{2}\rangle ^{128,126}}{\delta\langle r^{2}\rangle ^{126,124}}, 
\label{eqn:m2N}
\end{eqnarray}
introduced in Ref.\ \cite{Barzakh2018,DayGoodacre2016}. It provides a quantitative measure of the change of the slope (i.e. $\delta \left < r^2 \right>^{N,126}/\delta N$) of differential charge radii as a function of neutron number $N$ at $N=126$. The applicability of $\xi_{even}$ is restricted by the limited availability of the data for the $N$=128 isotones because for $84\leq Z \leq88$ they have half-lives ($t_{1/2}) <$300~\textmu s~\cite{Audi2017} which limits the potential for laser spectroscopy measurements. Thus, in these cases the $\xi_{even}$ indicator is replaced by 
\begin{eqnarray}
\xi^*_{even}=\frac{2}{N_0-126} \,\,\frac{\delta\langle r^{2}\rangle ^{N_0,126}}{\delta\langle r^{2}\rangle ^{126,124}}, 
\label{eqn:m2N*}
\end{eqnarray}
where $N_0$ is the lowest even neutron number at $N > 126$ with measured isotopic shift: $N_0 = 132$ for 
$_{84}$Po \cite{KBHKMR.91,Cocolios2011},
$_{85}$At \cite{Barzakh19}, 
$_{86}$Rn \cite{Otten1989,Borchers1987},
$_{87}$Fr \cite{Budin.14,Coc85,Dzuba.05}, 
$_{88}$Ra \cite{Wansbeek.12},
and $N_0 = 138$ for $^{89}$Ac~\cite{Ferrer2017}. 
For the $_{82}$Pb and $_{83}$Bi isotopes, the data for the nuclei with $N = 124, 126, 128$ were taken from \cite{Anselment1986} and \cite{Barzakh2018}, respectively. Note that for the Bi and Pb isotopes, for which the  $N = 128$ data are available, $\frac{2}{N_0-126} \,\, \delta\langle r^{2}\rangle ^{N_0,126} \approx \delta\langle r^{2}\rangle ^{128,126}$ ($N_{0}=130$ for Bi and $N_{0}=132$ for Pb) since differential radii increase nearly linearly for the $N>126$ nuclei under study.

The third is the three-point  indicator 
\begin{eqnarray}
\Delta \langle r^2 \rangle^{(3)}(N) && = \nonumber \\
= && \frac{1}{2} \left[ \langle r^2 \rangle(N-1) +
\langle r^2 \rangle(N+1) - 2  \langle r^2 \rangle(N) \right] = \nonumber \\
= && \frac{1}{2} \left[ r_{ch}^2(N-1) + r_{ch}^2(N+1) - 2  r_{ch}^2(N) \right] 
% \nonumber \\
\label{Delta-3}
\end{eqnarray}
which quantifies OES in charge radii.

%%%%%%%%%%%%%%%%%%%%%%%%%%%%%%%%%
\subsection{The kink in charge radii and its relation to underlying 
single-particle structure and pairing}\label{ch:discussion_kink}
%%%%%%%%%%%%%%%%%%%%%%%%%%%%%%%%%

%%%%%%%%%%%%%%%%%%%%%%%%%%%%%%%%%%%%%%%%%%%%%%
\begin{figure}
\includegraphics[width=1\linewidth]{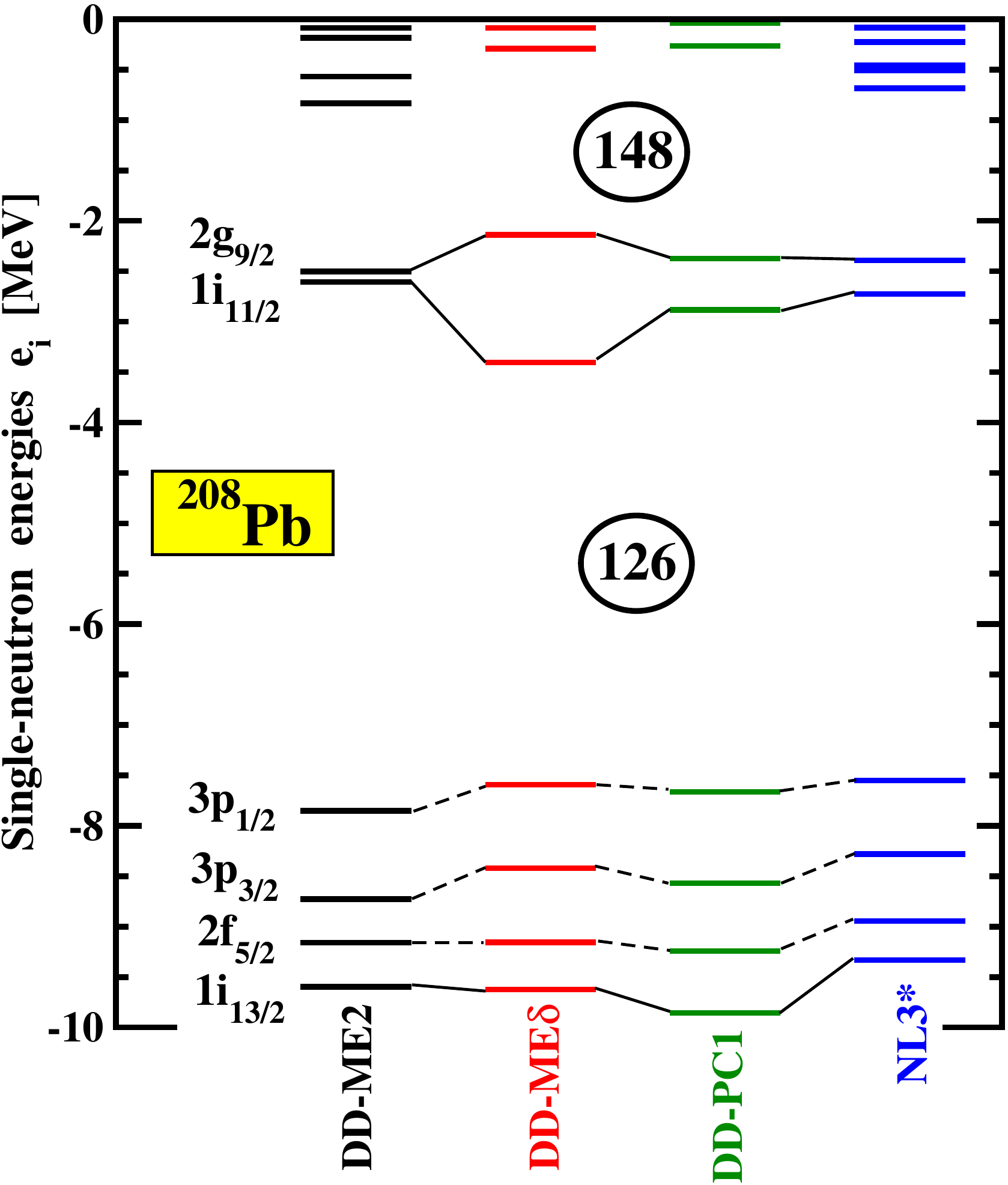}
\caption{\footnotesize The energies of neutron single-particle 
	states at spherical shape in $^{208}$Pb obtained in the calculations without pairing 
	with the indicated CEDFs. Solid and dashed connecting lines are used for positive- 
	and negative-parity states. Spherical gaps are indicated.}
\label{fig:pb208-sp}%
\end{figure}%
%%%%%%%%%%%%%%%%%%%%%%%%%%%%%%%%%%%%%%%%%%%%%%%

The differential charge radii  of the Pb and Hg isotopes are shown in Fig.\ \ref{fig:Pb-Hg-diff-charge-radii}. One can see that all of the employed CEDFs generate a kink at $N=126$ and that the best description is provided by the CEDF DD-ME2. Thus, it is important to understand which  physical features determine the differences between the functionals. For that we look at the energies of the neutron single-particle states and their occupation probabilities.

The energies of neutron single-particle states obtained in the $^{208}$Pb nucleus with the employed CEDFs are shown in Fig.~\ref{fig:pb208-sp}. One can see close similarities in the 
predictions of the energies and relative positions of the $2f_{5/2}$, $3p_{3/2}$ and $3p_{1/2}$ states occupied in the $N\leq 126$ nuclei. 
%On the other hand,
In contrast, the differences are more pronounced for the $1i_{11/2}$ and $2g_{9/2}$ orbitals located above the  $N=126$ shell closure. In all functionals, the $1i_{11/2}$ orbital is the lowest in energy\footnote{The order of these two orbitals is inverted in the majority of non-relativistic functionals (see Refs.\ \cite{Reinhard1995,Goddard2013}) and in many of them  this creates  a problem in the description of the kink in charge radii at $N=126$.} but the energy gap between these two orbitals strongly
depends on the functional. It is smallest in DD-ME2, gets larger in NL3* and DD-PC1 and become extremely large in DD-ME$\delta$. Because of this feature, and the fact that the kink in charge radii
at $N=126$ is defined by the interplay of the occupation of these two orbitals, we focus our discussion on the $N>126$ nuclei.

Let us analyze the slope of differential radii defined as $\delta \left < r^2 \right>^{N,N'}/\delta N$.
Note that in such an analysis we consider only even-even nuclei. The results of the calculations without 
pairing indicate that this slope is almost the same for $N<126$\footnote{The averaged slopes for the 
$N<126$ nuclei are almost the same in the calculations with and without pairing (see Ref.\ \cite{UAR.21} 
for detail).}  and $N>126$ nuclei when only the $2g_{9/2}$ states are occupied above $N=126$
(see Fig.\ \ref{fig:Pb-Hg-diff-charge-radii}(a) for Pb isotopes and Fig.\ \ref{fig:Pb-Hg-diff-charge-radii}(b)
for Hg isotopes). As a consequence, there would be either no kink or a 
very small kink in the charge radii at $N=126$. However, the situation drastically changes when only 
$1i_{11/2}$ states are occupied above $N=126$. This leads to a substantial increase
of the $\delta \left < r^2 \right>^{N,N'}/\delta N$ slope and as a result to a creation of large kink in the charge 
radii at $N=126$. This is related to the fact that the neutron $1i_{11/2}$ orbital is the 
$n=1$  orbital which overlaps more strongly with the majority of the proton orbitals than the $n=2$ 
neutron $2g_{9/2}$ orbital \cite{Goddard2013}. As a consequence, it provides a larger pull of the 
$1i_{11/2}$ neutron states  on  proton orbitals via the symmetry energy. 
This is despite the fact that in $^{208}$Pb the rms radius of the neutron $1i_{11/2}$ 
orbital (for example, $r_{ch}=6.4131$~fm in DD-ME2) is smaller than that of the neutron $2g_{9/2}$ one (for example, 
$r_{ch}=7.0227$~fm in DD-ME2) by $\approx 0.6$~fm in all employed CEDFs.

%%%%%%%%%%%%%%%%%%%%%%%%%%%%%%%%%%%%%%%%%%%%%%%%%
\begin{figure*}
\centering
\includegraphics[width=14.0cm]{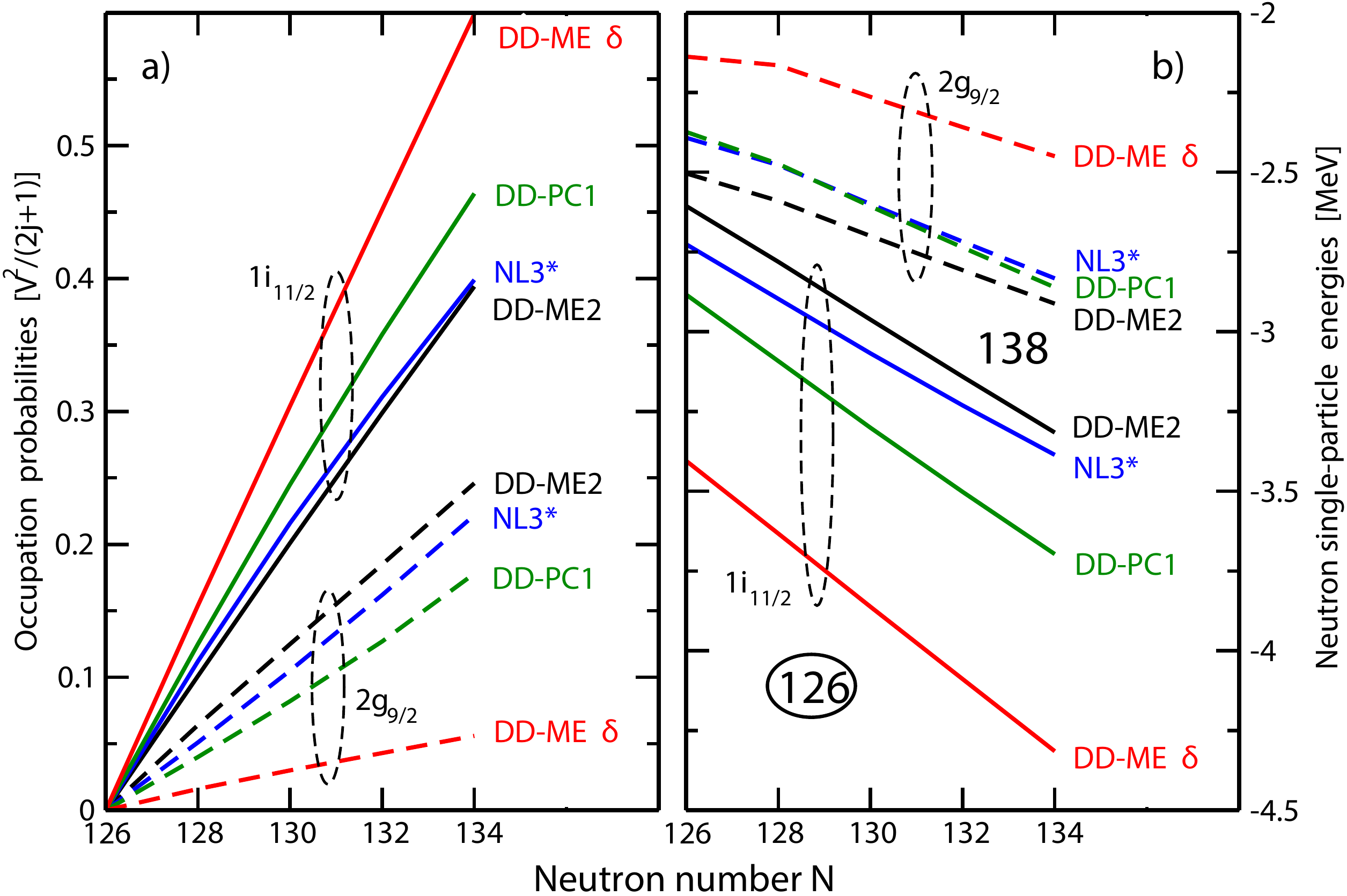}
\caption{\footnotesize (a) The evolution of the cumulative occupation probabilities $\text{v}^{2}_{state}$ of the neutron $2g_{9/2}$ and $1i_{11/2}$ orbitals as a function of neutron number in the $N\geq 126$ Pb nuclei for the indicated CEDFs. (b) The evolution of the energies of these  single-particle states as a function of neutron number. Note that both types of physical quantities are calculated in canonical basis. The neutron shell closure at $N=126$ and the energy gap between the $2g_{9/2}$ and $1i_{11/2}$ orbitals at $N=138$ are indicated. Note that this figure is based on the results of the calculations of even-even nuclei. Dashed ellipses envelope the states with the same structure.
}
\label{Pb-vv2-sp-energies}
\end{figure*}
%%%%%%%%%%%%%%%%%%%%%%%%%%%%%%%%%%%%%%%%

The inclusion of pairing modifies the situation in the $N>126$ nuclei in such a way that both of these orbitals become partially occupied (see Fig.~\ref{Pb-vv2-sp-energies}(a)). Note that we consider a cumulative occupation probability ${\rm v}^2_{state}$ which 
provides information on the filling of a given $j$-subshell: ${\rm v}^2_{state}$ could take any value
between 0 (unoccupied subshell) and $2j+1$ (fully occupied $j$-subshell). Pairing also leads to a partial occupation of the single-particle states located above the $N=148$ gap (see Fig.~\ref{fig:pb208-sp}), but their occupation probabilities are relatively  small because of the presence of this gap. Thus, for the sake of simplicity we focus our discussion on the interplay of the occupation of the $2g_{9/2}$ and $1i_{11/2}$ states and the consequences of this interplay on the $\delta \left< r^2 \right>^{N,N'}/\delta N$ slope. 

The large energy gap of 1.27 MeV between the $2g_{9/2}$ and $1i_{11/2}$ states in the DD-ME$\delta$ functional (see Fig.\ \ref{fig:pb208-sp}) is responsible for a predominant occupation of the lower lying $1i_{11/2}$ states (see  Fig.\ \ref{Pb-vv2-sp-energies}(a)). Considering the $N=134$ nucleus as an example, the eight neutrons outside of the $N=126$
shell closure are located almost entirely in the $1i_{11/2}$ subshell (${\rm v}^2_{1i_{11/2}}\approx 7.1$) with only a small portion  occupying $2g_{9/2}$ (${\rm v}^2_{2g_{9/2}}\approx 0.5$). This leads to a large $\delta \left < r^2 \right>^{N,N'}/\delta N$ slope (see Fig.\ \ref{fig:Pb-Hg-diff-charge-radii}). Note that this slope is the largest among the considered functionals and it is not far away from the one obtained in the calculations without pairing, when only the $1i_{11/2}$ states are occupied in the nuclei with $N>126$ (see Fig.~\ref{fig:Pb-Hg-diff-charge-radii}). The reduction of the gap between the $2g_{9/2}$ and $1i_{11/2}$ orbitals in the DD-PC1, NL3* and, especially, DD-ME2 functionals (see Figs.~\ref{fig:pb208-sp} and~\ref{Pb-vv2-sp-energies}(b)) is responsible for a decrease of the difference in the occupation of these orbitals (see Fig.~\ref{Pb-vv2-sp-energies}). For these three functionals, the eight neutrons outside of the $N=126$ shell closure in the $N=134$ nucleus are still located predominantly in the $1i_{11/2}$ subshell (${\rm v}^2_{1i_{11/2}}\approx 4.7$) but with a significant portion also found in $2g_{9/2}$ (${\rm v}^2_{2g_{9/2}}\approx 2.5$). This leads a reduction of the $\delta \left< r^2 \right>^{N,N'}/\delta N$ slope as compared with the case of the DD-ME$\delta$ functional. One can see in Fig.~\ref{fig:Pb-Hg-diff-charge-radii} that the experimental $\delta \left < r^2 \right>^{N,N'}/\delta N$ slope in the $N>126$ nuclei is reproduced with comparable accuracy by the DD-ME2, NL3* and DD-PC1 functionals. Note that with increasing neutron number $N$ the single-particle states become more bound, but for a given functional, the relative energies of the $2g_{9/2}$ and $1i_{11/2}$ states change only slightly (see Fig.~\ref{Pb-vv2-sp-energies}(b)). As a consequence, the occupation probabilities of the single-particle states behave as almost linear functions of neutron number (see Fig.~\ref{Pb-vv2-sp-energies}(a)).

\begin{figure}
	\includegraphics[width=1\linewidth]{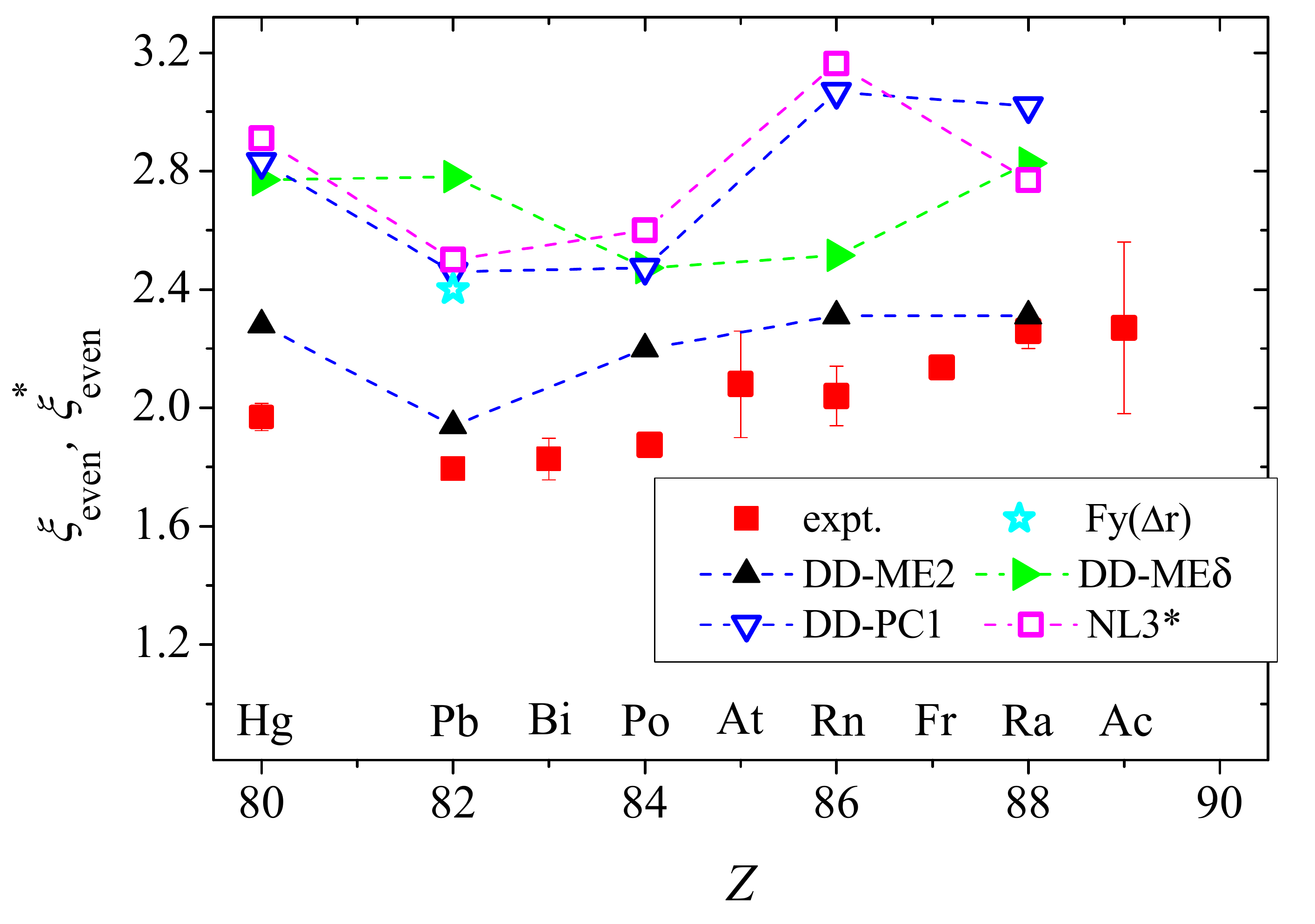}
	\caption {\footnotesize Experimental and calculated $\xi_{even}$ (for $Z\leq 83$)	and $\xi^*_{even}$ (for $Z\geq 84$) indicators. Note that the calculated results are presented only for even-even nuclei. The experimental values are determined using Eq.~(\ref{eqn:m2N}) and Eq.~(\ref{eqn:m2N*}), using data from~\cite{Barzakh2018, Goodacre-2021, Ulm1986, Anselment1986} and the references listed in the explanation of Eq.~(\ref{eqn:m2N*}), while the calculated ones are from the charge radii defined in the present paper and in Ref.\ \cite{Agbemava2014} which is publicly available at Ref.~\cite{Mass-Explorer}. The $\xi_{even}$(Pb) value for Fayans Fy$(\Delta r)$ functional was extracted from the data presented in Ref.~\cite{Gorges2019}). 
	} 
	\label{fig:dzeta-th-exp}
\end{figure}

The magnitude of the kink in charge radii at $N=126$ is better quantified by the $\xi_{even}$ and $\xi^*_{even}$ indicators defined in Eqs.\ (\ref{eqn:m2N}) and (\ref{eqn:m2N*}), respectively.  Experimentally available values of these indicators (both for even-even and odd nuclei) and calculated (only for even-even nuclei) are compared in Fig.~\ref{fig:dzeta-th-exp}. Mercury is the first element below $Z$=82 and only the second even-$Z$ element for which $\xi_{even}$  is experimentally determined; the kink in the mercury charge radii is of a similar magnitude to that of lead. Only experimental $\xi^*_{even}$ indicators are available for $Z>83$ nuclei. Note that experimental $\xi_{even}$ and $\xi^*_{even}$ indicators form a smooth curve which indicates that neither addition nor subtraction of proton(s) from the $Z=82$ nuclei affects drastically a kink in charge radii at $N=126$. Figure~\ref{fig:dzeta-th-exp} clearly shows that the best reproduction of this trend is achieved by the CEDF DD-ME2. Other functionals (including the Fayans Fy($\Delta r$) functional) deviate somewhat from experimental data.

This analysis clearly indicates that the evolution of charge radii with neutron and proton numbers is sensitive to the details of underlying singe-particle structure and the occupation probabilities of these states. Note that the latter depends on the type of pairing interaction employed in the calculations and its strength. It also indicates that the magnitude of the kink in charge radii at $N=126$ depends on the relative balance of the  occupation of the $2g_{9/2}$ and $1i_{11/2}$ orbitals. In general, the substantial occupation of the $1i_{11/2}$ orbital and the kink can be obtained even when the $2g_{9/2}$ orbital is located lower in energy than the $1i_{11/2}$ orbital (but still within its vicinity).This, for example, takes place in the non-relativistic HFB (NR-HFB) calculations with the semirealistic M3Y-P6a interaction, the spin-orbit properties of which were modified \cite{Nakada2015}  to improve the description of the charge radii of proton-magic nuclei \cite{Nakada2019,Nakada2015,Nakada2015a}. However, these calculations  underestimate the kink. A similar situation exists in the Skyrme DFT calculations with the SLy4mod functional presented in Ref.~\cite{Goddard2013}.

%%%%%%%%%%%%%%%%%%%%%%%%%%%%%%%%%%%%%%%%%%%%%%%
%%%%%%%%%%%%%%%%%%%%%%%%%%%%%%%%%%%%%%%%%%%%%%%

%%%%%%%%%%%%%%%%%%%%%%%%%%%%%%%%%%%
\subsection{Odd-even staggering in charge radii}
%%%%%%%%%%%%%%%%%%%%%%%%%%%%%%%%%%%

\begin{figure*}
	\includegraphics[width=16.0cm]{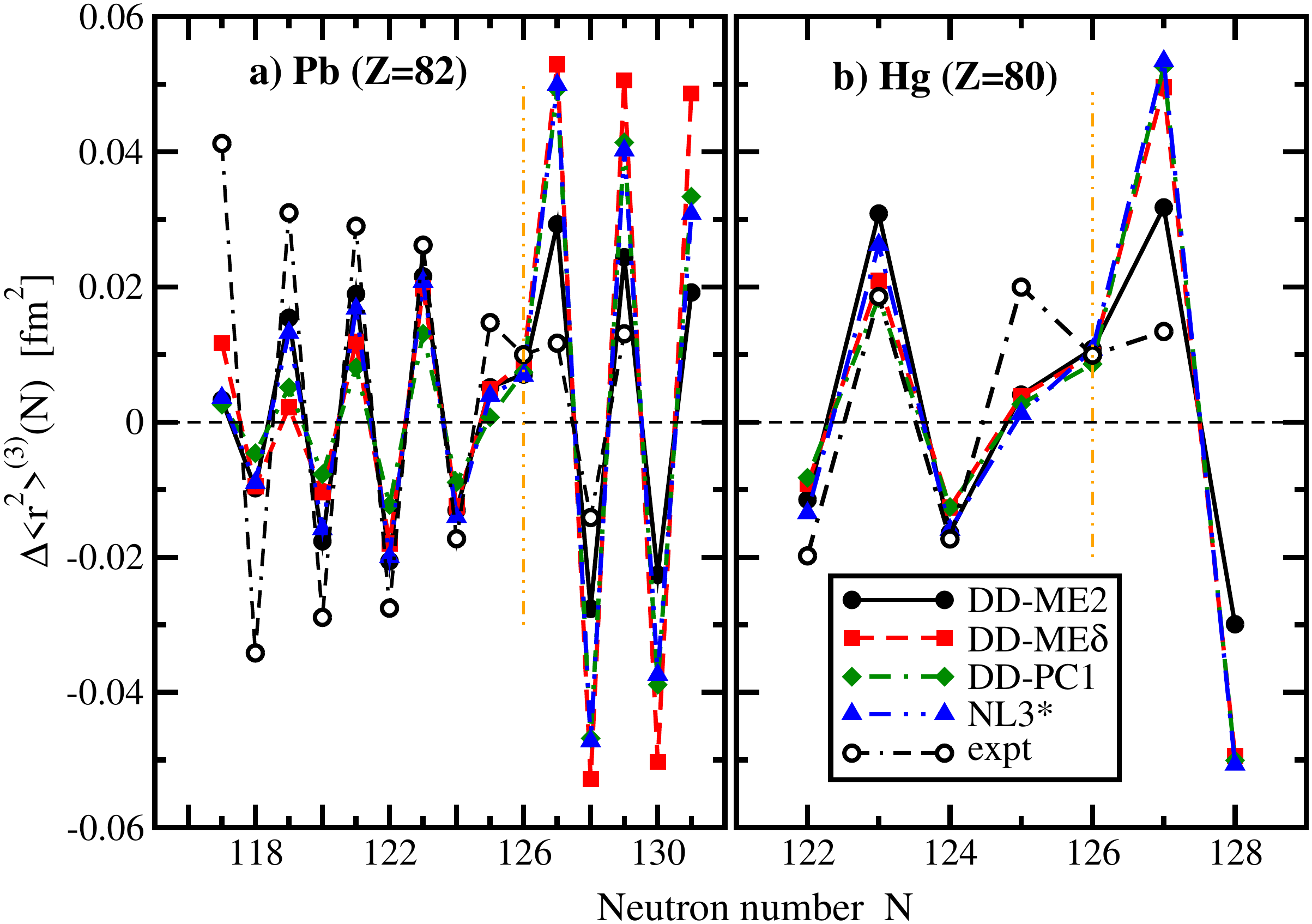}
	\caption{
		\footnotesize Comparison of experimental and theoretical $\Delta \left<r^2\right> ^{(3)} (N)$ values 
		for isotopes of lead (a) and mercury (b) , respectively. The EGS procedure is employed
		in odd-$A$ nuclei. Experimental data are taken from 
		Refs.\ \cite{Ulm1986,Anselment1986a,Goodacre-2021}. Vertical orange dot-dot-dashed 
		lines indicate $N=126$. 
	}

\label{fig:Pb-Hg-OES}
\end{figure*}

It was shown in Ref.~\cite{Goodacre-2021} both in the RHB calculations with DD-ME2 and in the NR-HFB studies with semi-realistic M3Y-P6a interaction that OES in charge radii is best reproduced when the EGS procedure is applied in odd-$A$ nuclei. In contrast, the experimental OES is significantly underestimated  when the LES procedure is used in odd-$A$ nuclei in the RHB framework for all nuclei under study and for $N < 126$ nuclei in the NR-HFB approach. The same behavior is observed also in the results of RHB calculations with CEDFs DD-ME$\delta$, DD-PC1 and NL3*, the results of which are shown in Fig.~\ref{fig:Pb-Hg-OES}. This figure shows also that there is some dependence of the magnitude of the $\Delta \left<r^2\right> ^{(3)} (N)$ values on the employed functional which comes from the differences in the energies of the single-particle states and their occupations (see Figs.~\ref{fig:pb208-sp} and~\ref{Pb-vv2-sp-energies}).

%%%%%%%%%%%%%%%%%%%%%%%%%%%%%%%%%%%%%%%%%%%%%%
%%%%%%%%%%%%%%%%%%%%%%%%%%%%%%%%%%%%%%%%%%%%%%%

Particle-vibration coupling (PVC) plays a critical role in the emergence of such significant OES in charge radii because it leads to the nucleonic configuration with the blocked state corresponding to the experimental ground state (see Ref.~\cite{Goodacre-2021} for details). Let us illustrate that with the case of the $N>126$ Pb nuclei. In the odd-$A$ isotopes, the PVC coupling lowers the $2g_{9/2}$ state below the $1i_{11/2}$ state making it the ground state despite the fact that at the mean field level (as well as in even-even nuclei) the energy of the $2g_{9/2}$ state is higher than that of the $1i_{11/2}$ state (see Fig.~5 of Ref.~\cite{Litvinova2011}). Although these results were obtained with NL3*, a comparable effect of PVC on relative energies of these states is expected in other functionals. However, the feasibility of such a scenario also crucially depends on the relative energies of the single-particle orbitals of interest at the mean field level. For example, if the energy gap between the $2g_{9/2}$ and $1i_{11/2}$ states is too large (like in the case of the DD-ME$\delta$ functional, see Figs.~\ref{fig:pb208-sp} and \ref{Pb-vv2-sp-energies}(b)), the impact of the PVC would most likely not be enough  to make the $2g_{9/2}$ state as a ground state. This functional however suffers from some significant deficiencies in the $Z>82$ region which probably are the consequences of the overly large energy gap between these two states. For example, it does not predict octupole deformed actinides~\cite{AAR.16} and predicts fission barriers in superheavy nuclei which are too small to make them relatively stable~\cite{AARR.17}. One has to keep in mind, however, that the interaction DD-ME$\delta$ is different from the other interactions discussed here. It is the most microscopic one among considered functionals: only four parameters at the saturation density are fitted to finite nuclei and the full density dependence of the parameters is derived from ab-initio calculations. On the contrary, the other interactions contain an additional 2 (NL3*), 4 (DD-ME2) or 6 (DD-PC1) phenomenological parameters for the fine-tuning of the  density dependence.

%%%%%%%%%%%%%%%%%%%%%%%%%%%%%%%%%%%
\subsection{Binding energies and $\Delta^{(3)}_E$ indicators}
%\subsection{Binding energies and mass filters}

%%%%%%%%%%%%%%%%%%%%%%%%%%%%%%%%%%%

\begin{figure}
\includegraphics[width=1\linewidth]{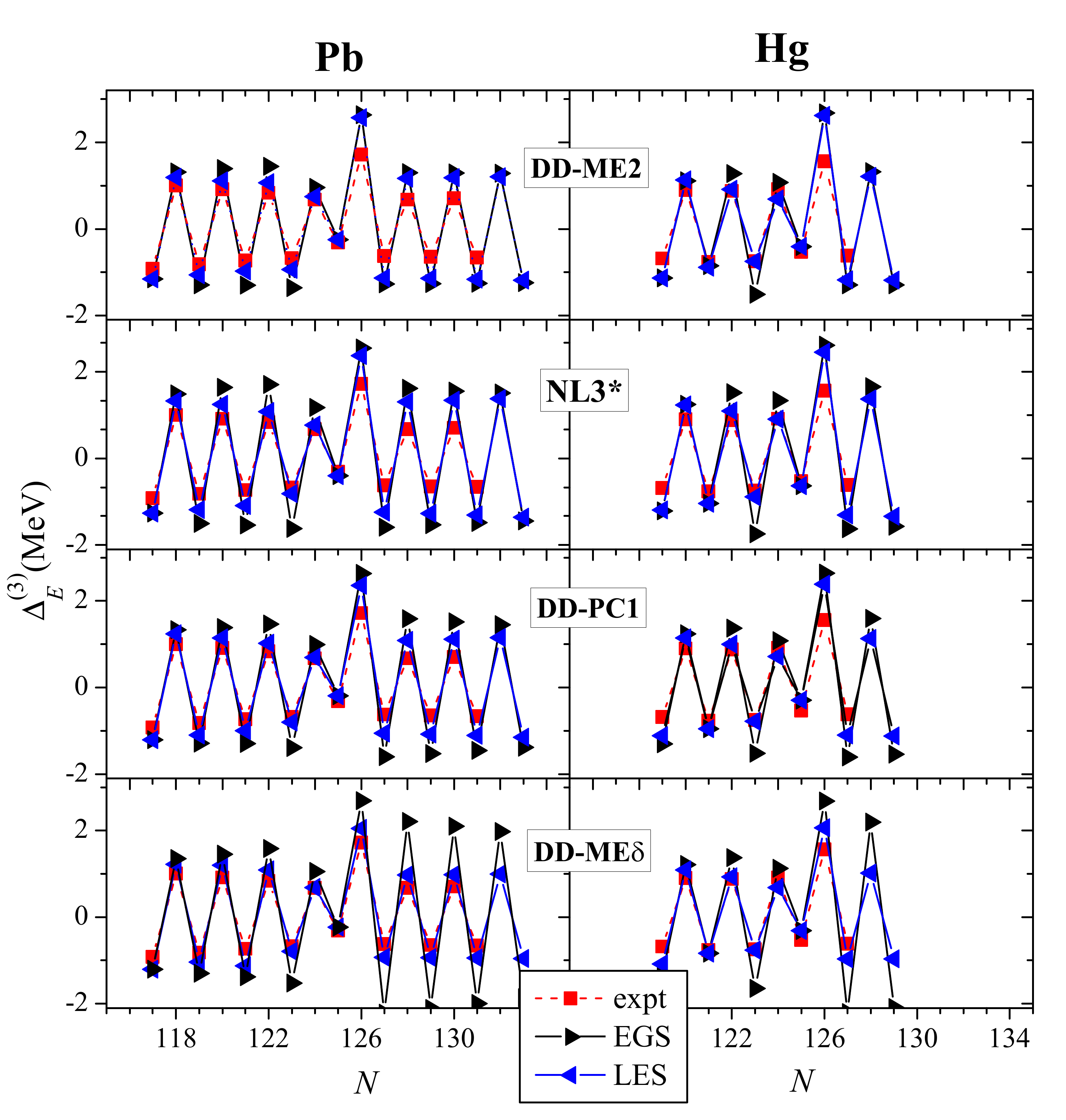}
\caption{
	\footnotesize Experimental and calculated odd-even staggerings in binding energies $\Delta^{(3)}_E$ of the Pb and Hg isotopes under study. The calculated results are defined for the EGS and	LES procedures in odd-$A$ nuclei.  
} 
\label{fig:D3-energy-th-exp}
\end{figure}

Empirical approaches considering binding energies can help with understanding the interplay between nucleons. With exception of the DD-ME$\delta$ functional, there is a good  agreement between the calculated and the experimental binding energies for both lead and mercury  isotopes. For $^{198-214}$Pb and $^{201-208}$Hg nuclei, the rms deviations from experiment are 1.3~MeV and 0.6~MeV for NL3* CEDF, 1.3~MeV and 1.1~MeV for DD-ME2,  1.8~MeV and 2.1~MeV for DD-PC1, and 4.0~MeV and 3.9~MeV for DD-ME$\delta$, respectively. To highlight the odd-even staggering of binding energies along the isotopic chains, an indicator in the form of 
\begin{align}
\Delta^{(3)}_E(Z,N) = \frac{1}{2}\left[ B(Z,N-1) -2B(Z,N) + B(Z,N+1)\right]
\label{eq:three-point-estimator}
\end{align}
was used. Where $B(N,Z)$ is the binding energy for a nucleus with proton number $Z$ and neutron number $N$. The odd-even staggerings in the binding energies\footnote{This quantity  is frequently used as a pairing indicator. However, in no way it should be considered as a clean measure of pairing correlations since it is polluted by time-odd mean fields and particle-vibration coupling in odd-$A$ nuclei (see detailed discussion in Ref.~\cite{TA.21}).} $\Delta^{(3)}_E$ are reproduced reasonably well (see Fig.\ \ref{fig:D3-energy-th-exp}), especially when the LES are used in odd-$A$ nuclei. The level of agreement is comparable with that provided by the Fayans Fy($\Delta r$) functionals in Ref.~\cite{Reinhard2017}.  

Figure~\ref{fig:D3-energy-th-exp} also illustrates the challenges faced by all existing theories. Polarization effects in deformation/radial density distributions and pairing depend on blocked state in odd-$A$ nuclei. Thus, on the one hand, the $\Delta^{(3)}_E$ indicators are distorted by incorrect polarizations effects when a wrong (as compared with experiment) state is used for the ground state of odd-$A$ nucleus. The related uncertainties in binding energies due to polarization effects in the pairing channel  are on the  level of 150~keV and those due to deformation/radial density distribution polarization effects are more difficult to estimate but are expected typically to be less than 100~keV. On the other hand, large theoretical uncertainties in the predictions of single-particle energies (see Refs.~\cite{Litvinova2011,Afanasjev2015,AS.11} and Fig.~\ref{fig:pb208-sp}) reveal themselves when the $\Delta^{(3)}_E$ indicators are defined using the EGS procedure in odd-$A$ nuclei. This is especially pronounced in the calculations for the $N>126$ nuclei with the DD-ME$\delta$ functional (see Fig.\ \ref{fig:D3-energy-th-exp}) which is characterized by a large energy gap between the $2g_{9/2}$ and $1i_{11/2}$ subshells (see Fig.~\ref{fig:pb208-sp}). Comparing these uncertainties, it is safer to use the LES procedure in the definition of the $\Delta^{(3)}_E$ indicators and their association with pairing indicators (see detailed discussion in Ref.\ \cite{TA.21}). Note that the calculations somewhat overestimate experimental $\Delta^{(3)}_E$ indicators. The PVC provides additional binding (on the level of few 100 keV) to the ground states of odd-$A$ nuclei. Thus, its inclusion into the calculations is expected to decrease the calculated $\Delta^{(3)}_E$ and, as a consequence, to improve the description of experimental data.

%%%%%%%%%%%%%%%%%%%%%%%%%%%%%%%%%%%%%%%%%%%%%%%
%%%%%%%%%%%%%%%%%%%%%%%%%%%%%%%%%%%%%%%%%%%%%%%

%%%%%%%%%%%%%%%%%%
\section{Conclusion}
\label{conclusion}
%%%%%%%%%%%%%%%%%%

The kink in the $\delta \langle r^2\rangle$ systematics of the mercury isotopic chain has been analyzed considering a variety of dimensionless parameters and employed in the comparison of the results of a range of covariant energy density
functionals. The results of mass measurements of $^{206-208}$Hg are presented which improve upon the precision of previously measured values. The observed $Z$ dependence of the $g$-factors of $I$=9/2, $N$=127 isotones is interpreted as the result of CP2 corrections. The magnitude of the extracted neutron effective charge for $^{207}$Hg suggests a rapidly increasing quadrupole core polarization when moving away from the $Z$=82 proton shell.   

The theoretical analysis of charge radii and related indicators in the Pb and Hg isotopic chain has been performed within the RHB framework using several CEDFs characterized by different single-particle properties. This analysis supports the conclusion that the kink at $N=126$ in $\delta \left<r^2\right>^{N,N'}$ originates from the occupation of the $\nu 1 i_{11/2}$ orbital located above the $N=126$ shell gap. The pairing effect does not play a critical role here since the kink is present also in the calculations without pairing. This is in contrast to non-relativistic Skyrme and Fayans functionals, in which the pairing becomes a dominant contributor to the kink and OES~\cite{Gorges2019}.  However, the pairing plays an important role in defining the magnitude of the kink which depends on the balance of the occupation of the $1i_{11/2}$ and $2g_{9/2}$ orbitals. This balance sensitively depends on the relative energies of these two orbitals. The DD-ME2 functional provides the best description of kinks at $N=126$ not only in the Pb and Hg isotopes but also in all isotopic chains for which experimental data is available. A reasonable description of OES in charge radii has been achieved with all employed functionals. This confirms a new mechanism of OES suggested in Ref.\ \cite{Goodacre-2021} which is related to the staggering in the occupation of neutron orbitals between odd and even isotopes facilitated by PVC in odd-mass nuclei.

\section{Acknowledgements} 
\label{sec:Acknowledgements}This project has received funding through the European Union's Seventh Framework Programme for Research and Technological Development under grant agreements 267194 (COFUND), 262010 (ENSAR), 289191 (LA$^{3}$NET) and 267216 (PEGASUS). This project  has received funding from the European Union’s Horizon 2020 research and innovation programme grant agreement No 654002 (ENSAR2). This material is based upon work supported by the US Department of Energy, Office of Science, Office of Nuclear Physics under Award No. DE-SC0013037, and by the Deutsche Forschungsgemeinschaft (DFG, German Research Foundation) under under Germany’s Excellence Strategy EXC-2094-390783311, ORIGINS. S.S. acknowledges a SB PhD grant from the former Belgian Agency for Innovation by Science and Technology (IWT), now incorporated in FWO-Vlaanderen. This work was supported by the RFBR according to the research project N 19-02-00005; the ERC Consolidator Grant No.~648381; the IUAP-Belgian State Science Policy (BRIX network P7/12), FWO-Vlaanderen (Belgium) and GOA’s 10/010 and 10/05 and starting grant STG 15/031 from KU Leuven; the Science and Technology Facilities Council Consolidated Grant Nos. ST/F012071/1 and ST/P003885/1, Continuation Grant No.~ST/J000159/1 and Ernest Rutherford Grant No.~ST/L002868/1; the Slovak Research and Development Agency, contract No. APVV-14-0524; the French IN2P3, the BMBF (German Federal Ministry for Education and Research) grants Nos. 05P12HGCI1, 05P15HGCIA, 05E18CHA, and 05P18HGCIA.

\appendix
\section{Nuclear observables in the hyperfine structure}\label{ch:app:NO_HFS}
Nuclear observables were extracted from the experimentally measured hfs through the application of standard methods~\cite{Otten1989}. The substate weighted centroid $\nu _0$ and the hyperfine $a$ and $b$ parameters were extracted from the fitted spectra, where the shift $\Delta \nu ^F$ of the state $F$ of the hyperfine multiplet from $\nu _0$ is given as

\begin{equation}
\Delta \nu ^F =a\frac{K}{2} + b \frac{\frac{3}{4} K(K+1) - I(I + 1)J(J + 1)}{2(2I - 1)(2J - 1)IJ},
\label{nuF}
\end{equation}

\noindent where $K$=$(F(F+1)-I(I+1)-J(J-1))$, $I$ is the nuclear spin and $J$ is the atomic angular momentum. The extraction of $\nu _0$ enabled the calculation of $\delta \nu ^{A,A'}$, the isotope shift between $A$ the isotope under investigation and $A'$ a reference isotope. $\delta\langle r^{2}\rangle^{A,A'}$, the change in the mean square charge radius of $A$ with respect $A'$ was extracted as

\begin{equation}
\delta \nu ^{A,A'} = F_{\lambda}K(Z)\delta\langle r^{2}\rangle^{A,A'} + M\times\frac{A-A'}{AA'},
\label{mscr}
\end{equation}

\noindent where for the spectroscopic transition $F_{254 nm}=-55.36$~GHz~fm$^{-2}$~\cite{Ulm1986}, $K(Z=80)=0.931$ (taking into account~\cite{Torbohm1985,Fricke}) is a calculable correction factor and $M$ is the mass shift factor representing the sum of the normal mass shift ($M_{NMS}$) and the specific mass shift ($M(Z=80)=(1\pm 0.5)\cdot M_{NMS}$~\cite{Ulm1986}).

The magnetic moments $\mu_A$ were calculated as
\begin{equation}
\mu _A = \mu_{ref} \times \dfrac{I_A}{I_{ref}} \times \dfrac{a_A}{a_{ref}} \times (1+^{ref}\Delta^A)
\end{equation}
\noindent where the isomer $^{199m}$Hg ($I=13/2$) was used as a reference nucleus with $\mu(^{199m}Hg)=-1.0147(8)~\mu_N$~\cite{Reimann1973} and $a(^{199m}Hg) = -2298.3(2)$~MHz~\cite{Reimann1973}. An additional correction for the hyperfine anomaly (HFA) is included for the $\mu$ values presented in Table~\ref{tab:Exp_results}. Moskowitz and Lombardi~\cite{Moskowitz1973} demonstrated that for mercury isotopes the “Bohr-Weisskopf” component ($^{A_1}\Delta ^{A_2}_{BW}$)~\cite{Bohr1950} is dominant, thus the “Breit-Rosenthal” component ($^{A_1}\Delta ^{A_2}_{BR}$)~\cite{Rosenthal1932} can be ignored. The following relation between the magnetic moments and the HFA, the so called Moskowitz-Lombardi (ML) rule was used, determining $^{A_1}\Delta ^{A_2}_{BW}$~\cite{Bohr1950} as
\begin{equation}
^{A_1}\Delta ^{A_2}_{BW}=\pm \alpha \times \Bigg( \dfrac{1}{\mu _1}-\dfrac{1}{\mu _2}\Bigg), I=l\pm \dfrac{1}{2},
\end{equation}
\noindent where $\alpha=1\times 10^{-2}$ and $l$ is the orbital momentum of the last neutron. The ML rule was verified later by the microscopic theory~\cite{Fujita1975}. The application of the ML rule to estimate the Bohr-Weisskopf correction of the magnetic moment of $^{207}$Hg is justified based on the successful reproduction of the experimental HFA by this rule for  neutron single particle states in mercury nuclei across a rather large range of masses~\cite{Moskowitz1973}.

For previously measured isotopes and isomers the maximum deviation of the experimental $^{A_1}\Delta^{199}_{BW}$ from the ML calculation is equal to 2$\times 10^{-3}$. Correspondingly, we conservatively estimated the error of the ML prediction for $^{207}\Delta^{199}_{BW}$ as 5$\times 10^{-3}$. The uncertainty of this correction was estimated based on the omitted  $^{A_1}\Delta ^{A_2}_{BR}$ correction. It was shown in~\cite{Martensson-Pendrill1995} that $^{A_1}\Delta^{A_2}_{BR}$ is proportional to $\delta \langle r^2\rangle^{A_1, A_2}$. Thus $^{207}\Delta^{199}_{BR}$ for $^{207}$Hg can be estimated by scaling the calculated $^{201}\Delta^{199}_{BR}$~\cite{Rosenberg1972}. It should be noted that $^{205}\Delta ^{203}_{BR}$ for thallium isotopes, calculated in~\cite{Rosenberg1972} by solving the one-electron Dirac equation, practically coincides with that calculated in~\cite{Martensson-Pendrill1995} by the Dirac-Fock approach. Taking into account the independence of the $^{A_1}\Delta ^{A_2}_{BR}$ correction on the details of the atomic calculations and the uncertainty of $\delta \langle r^2\rangle$, we estimate the uncertainty of this correction as 10\%.

The spectroscopic quadrupole moments $Q^A_s$ were calculated using the relation
\begin{equation}
Q^A_s=\dfrac{b_A}{b_{^{201}\text{Hg}}}Q^{^{201}\text{Hg}}_s,\\
\end{equation}
\noindent where $Q^{^{201}\text{Hg}}_s$=$0.387(6)$ b (from~\cite{Bieron2005}) and $b_{^{201}\text{Hg}}$=$-280.107(5)$~MHz~\cite{Kohler1961}. The results are presented in Table~\ref{tab:Exp_results}.

\section{Mass spectrometry methods}\label{ch:app:mass_measurements}

The mass measurements were performed with the ISOLTRAP mass-spectrometer \cite{Mukherjee2008}. As described in parts above, the setup consists of four ion traps for beam preparation and mass measurements. First, a linear radio-frequency quadrupole trap was used to accumulate, cool and bunch the quasi-continuous radioactive ion beam delivered by ISOLDE \cite{Herfurth2001}. Using a pulsed drift tube, the energy of the bunched beam is then reduced from the initial \SI{30}{\kilo\electronvolt} to \SI{3.2}{\kilo\electronvolt}.

Ions are captured by the Multi-Reflection Time-of-Flight Mass Spectrometer/Mass Separator (MR-ToF MS)~\cite{Wolf2013a} using the in-trap lift electrode \cite{Wolf2012b}, following this, they undergo a certain number of revolutions in-between the mirror electrodes of the device. Here, the time-of-flight is given as $t=a\sqrt{m}+b$, where $a$ and $b$ are device-dependent parameters. Due to the mass-dependence of the trapped ions moving at the same kinetic energy, the different isobaric species separate in time-of-flight. In mass-spectrometry mode, the MR-ToF MS can be used to determine the mass of the ion of interest by using well known reference masses to account for the calibration parameters $a$ and $b$. This is done by expressing the mass $m$ using the so-called $C_{ToF}$-value \cite{Wienholtz2013}:
\begin{align}
m^{1/2}=C_{ToF}\left(m_1^{1/2}-m_2^{1/2}\right)+\frac{1}{2}\left(m_1^{1/2}+m_2^{1/2}\right),
\label{eq:Ctof}
\end{align}
where $m_1, m_2$ and $t_1, t_2$ are the masses and time-of-flights of the two reference species, respectively, and $C_{ToF}=(2t-t_1-t_2)/\left[2(t_1-t_2)\right]$. An example of a time-of-flight spectrum is presented in Figs.~\ref{fig:ToF}(b) and \ref{fig:ToF}(c). %For the extracted $C_{ToF}$-values for the mass measurement of $^{208}$Hg, see Figure \ref{fig:208Hg_Ctof}.

In mass-separation mode, the MR-ToF MS selectively ejects the ions of interest towards the Penning traps located downstream of the setup. In the so-called preparation Penning trap, the ions are captured and cooled using a mass-selective buffer-gas method to improve beam emittance \cite{Savard1991} and to further reduce contamination. Subsequently, the ions are ejected and recaptured in the precision Penning trap in which the high precision mass measurement is performed. In the present measurement, this is achieved by employing the Time-of-Flight Ion Cyclotron Resonance technique (ToF-ICR), which determines the cyclotron frequency of the trapped ions by scanning the frequency of an applied quadrupolar electric field \cite{Konig1995}. This frequency can be written as $\omega_c=\frac{qB}{m}$, where $q$ is the electric charge of the ion, $B$ is the magnetic field, and $m$ is the ion mass. By performing cross-reference measurements with known mass calibrants, the mass $m$ of the ion of interest can be extracted by comparing its cyclotron frequency with that of a well-known mass. Expressed as a frequency ratio
\begin{align}
R = \frac{\omega_{c,ref}}{\omega_{c}}=\frac{m}{m_{ref}},
\label{eq:R_tof}
\end{align}
the magnetic field and the charge cancel out. For the atomic mass, the electron mass is added to the measured ion mass. The ionization energy is negligible. 

Expressing the ion masses in MR-ToF MS and ToF-ICR measurements via the $C_{ToF}$-value and the frequency ratio $R$, respectively, facilitates an easy recalculation of the mass of interest in case one of the employed reference masses are measured more precisely in the future.

\bibliography{Hg_PRC}

\end{document}